\documentclass[aps,twocolumn,superscriptaddress,amssymb,prb,floatfix,preprintnumbers]{revtex4-2}
\usepackage{graphicx,color}
\usepackage{dcolumn}
\usepackage{amsmath}
\usepackage{amssymb}

\def\be{\begin{equation}}
\def\ee{\end{equation}}

\begin{document}

\thispagestyle{myheadings}

\title{Understanding the onset of negative electronic compressibility in one- and two-band 2D electron gases: Application to LaAlO$_3$/SrTiO$_3$}

\author{A. D. Mahabir}
\affiliation{Department of Physics, University of North Florida, Jacksonville, FL 32224, USA}
\affiliation{Department of Physics, University of Connecticut, Storrs, CT 06269, USA}
\author{A. V. Balatsky}
\affiliation{Department of Physics, University of Connecticut, Storrs, CT 06269, USA}
\affiliation{NORDITA, Roslagstullsbacken 23, 106 91 Stockholm, Sweden}
\author{J. T. Haraldsen}
\affiliation{Department of Physics, University of North Florida, Jacksonville, FL 32224, USA}


\date{\today}

\begin{abstract}

{We investigate the effects of  two electronic bands at the negative electronic compressibility (NEC) in a two-dimensional electron gas (2DEG). We use a simple homogeneous model with Coulombic interactions and first-order multi-band coupling to examine the role of effective mass and relative permittivity in relation to the critical carrier density, where compressibility turns negative. We demonstrate that the population of a second band, along with the presence of inter-band coupling, can dramatically change the cross-over carrier density. Given the difficulty in determining and confirming multi-band electronic systems, this model provides a potential method for identifying multi-band electronic systems using precise bulk electronic properties measurements. To help illustrate this method, we apply our results to the observed NEC in the 2D electron gas at the interface of LaAlO$_3$/SrTiO$_3$ (LAO/STO) and determine that, for the known parameters of LAO/STO, the system is likely a realization of a two-band 2D electron gas. Furthermore, we provide general limits on the inter-band coupling with respect to the electronic band population.}


\end{abstract}

\maketitle

\section{Introduction}

Complex oxides heterostructures have provided an exciting platform for various physical phenomena for over two decades, demonstrating the fascinating interplay between competing states depending on composition, doping, substitution, and structure\cite{hebe:09,prel:05}.  Highly tunable electron-electron interactions at interfaces have been shown to host states with charge, spin, and orbital orderings\cite{khom:97,zubk:11,chak:07,lee:18,rein:12,hara:12} that emerge from the two-dimensional electron gas (2DEG) at the interface.\cite{kozu:11,park:10,popo:20,kiri:17,pesq:14,stem:14}

While there are many interesting properties of the 2DEG, we focus on the observations of negative electronic compressibility (NEC). NEC is produced through electron-electron interactions, where the exchange and Coulomb energies outweigh the overall kinetic energy of the electronic system\cite{bell:81,kopp:09,tana:89,scha:01,skin:10}. {Electronic compressibility is typically negative for dynamic, open 2D systems}\cite{bell:81,krav:90,eise:92,eise:94,dult:00,kusm:08,junq:19} in non-equilibrium states, although recently, 3D NEC systems have been discussed\cite{wen:20}.

The 2DEG state of the complex oxide heterostructure LaAlO$_3$/SrTiO$_3$ (LAO/STO) exhibits a NEC. Over the last decade, measurements of the quantum capacitance have shown that the LAO/STO interface displays a large NEC around a carrier density of 10$^{13}$ cm$^{-2}$.\cite{li:11,tink:12,smin:17}. Aside from NEC, LAO/STO interfaces were shown to host the possibly unconventional superconducting states at $T_{c}\sim0.2$ K\cite{ohto:04,reyr:07,bisc:12}, which has led to fundamental questions about the nature of the superconducting state and its relation to the nature of the normal state.\cite{cavi:08,bell:09,fern:13,hara:13,sing:18,bosc:15}. It is also known that superconductivity can be enhanced to 0.3-0.4 K by applying an electric field\cite{cavi:08,bell:09} and is observed to increase with strain\cite{dunn:18,herr:19}. Therefore, there is an exciting possibility that NEC and superconductivity might be related.

Observations point to the very robust superconducting interactions in STO that are controlled by doping and dimensionality. Literature has shown that the electron mobility observed at the interface is related to the electronic states within STO alone and is not dependent on the adjacent material\cite{koon:67,binn:80,fern:13,smin:17,veer:13,gudu:13,song:18,sing:18,bosc:15}. Furthermore, while still controversial, recent experiments have also found strong evidence that the maximum $T_{c}$ in the heterostructures is achieved once an extra electron band becomes occupied\cite{josh:12}. Currently, there are a few proposals that the superconductivity in LAO/STO may be a realization of a multi-band system\cite{fern:13,smin:17,veer:13,gudu:13,josh:12,song:18}, where band structure calculations have illustrated the role of split bands in electronic transport as a function of doping\cite{vanm:08}. {However, this is not without controversy, as some reports indicate that this not the case \cite{bisc:12,bosc:15,smin:18,sing:18}, while others the claim of multi-band electronic structure still need to be looked at carefully \cite{khom:97,zubk:11,chak:07}.}

The determination of a multi-band superconducting state is quite complicated as the multi-band phenomenon's conclusive evidence can take decades {due the difficulty in resolving the small energy gap between bands which is not well resolved  by most techniques.} However, not long after the theoretical basis\cite{suhl:59}, multi-band superconductivity was discovered in transition-metal superconductors, including Nb-doped STO\cite{rade:66,shen:65,binn:80}. In the case of Pb superconductors, the theory suggested that two-band superconductivity was possible in 1965\cite{benn:65}, but it was not experimentally confirmed until 2000\cite{shor:00}, despite earlier experimental evidence\cite{toml:75}. Detailed $ab$-$initio$ calculations supporting the multi-band scenario were performed in 2009\cite{bers:09}.  

This paper aims to examine the differences in electronic compressibility between one- and two-band models and hopes to provide clarity and guidance for understanding the nature of collective electronic states at 2D interfaces (i.e., multi-band superconductivity). Therefore, in this study, we examine the one and two-band models of a 2DEG and investigate the onset of negative compressibility. Using a homogeneous 2D electron model and a two-band description of the electron interactions, we examine the dependence of the negative compressibility crossover point with respect to the carrier density. In general, we find the increases in effective mass leads  to the rise in critical carrier density for NEC. In contrast, an increase in the relative permittivity of the 2DEG leads to a decrease in the critical carrier density. Furthermore, the addition of a second band lowers the overall {crossover} carrier density, where the presence of inter-band coupling will dramatically reduce the polarizability of the 2DEG. The analysis presented in this work establishes features in the bulk electronic properties that allow for {the possible} identification of multiple bands in 2DEG systems. Additionally, through the comparison with previous experimental results for LAO/STO\cite{li:11,tink:12,smin:17,song:18}, we find that the NEC of LAO/STO is consistent with a two-band model of the 2DEG, which provides further evidence that the LAO/STO interface electron states may be a realization of a multi-band system. Our findings also provide a straightforward method for the  identification of electron band populations and coupling between the bands.

The structure of the paper is as follows: after the Introduction, we discuss the general attributes of the negative compressibility state in Sec II. In Sec. III, we revisit the electron gas free energy calculation for the multi-band case. In Sec IV, we discuss our results in the context of LAO/STO interface and conclude in Sec. V with a summary of the results. 

\begin{figure}
\includegraphics[width=\linewidth]{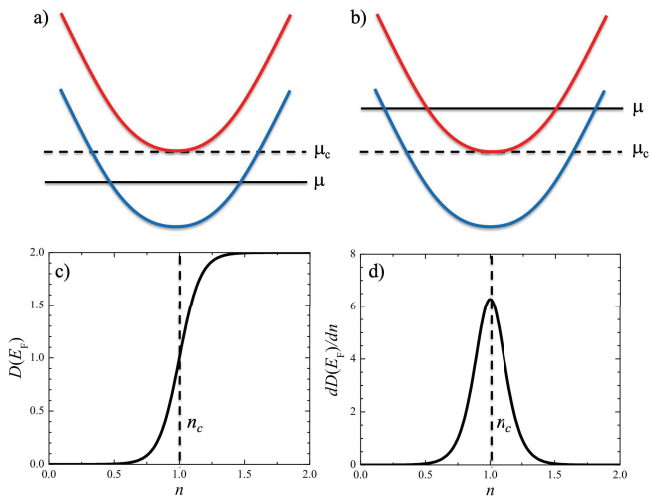}
\caption{(Color Online) Illustration of the two-bands with the (a) $\mu < \mu_c$ and (b) $\mu > \mu_c$. (c) The DOS as a function of $n$ for the two-band system. (d) $dD(E_{{\rm F}})/dn$ as function of $n$. }
\label{crit}
\end{figure}

\section{Electronic Compressibility}

Electronic compressibility, the response of the chemical potential to changes in the carrier density, is given by
\be
\kappa = \left(n^2 \frac{d\mu}{dn}\right)^{-1}
\ee
where $\mu$ is the chemical potential and $n$ is the carrier density.  Typically, an increase in the carrier density produces a positive increase of the chemical potential\cite{krav:90,tink:12}. Because of standard thermodynamic constraints, electronic compressibility tends to be positive. However, if the system is interacting, in some cases, the compressibility shift is negative, resulting in the so called negative electronic compressibility\cite{wall:72}. NEC is typically found in 2D materials due to topological effects\cite{bell:81,krav:90,eise:92,dult:00,kusm:08}. Recently it has also been suggested in 3D materials\cite{wen:20}.

If we assume a two-band model with simple parabolic dispersion (as shown in Fig. \ref{crit}), then as the chemical potential increases to $\mu/\mu_c$ = 1, the second band becomes populated and the carrier density will increase.  At a  critical density, one will see a dramatic increase in the Density of States (DOS) $D(E_{{\rm F}})$ (shown in Fig. \ref{crit}(c)) as a result of second band being populated.

We start with inverse compressibility $\kappa^{-1}$  being proportional to $d\mu/dn$, and can be written as

\be
\frac{d\mu}{dn} = \frac{d\mu}{dD(E_{{\rm F}})} \frac{dD(E_{{\rm F}})}{dn}.
\ee
where $dD(E_f)/dn$ is essentially $\delta(D(E_{{\rm F}})-n)$. Given a standard experimental width, $dD(E_{{\rm F}})/dn$ will resemble Fig. \ref{crit}(d). 

For a two dimensional electron gas (2DEG) at $T$ = 0, the chemical potential is equivalent to the Fermi Energy $E_{{\rm F}}$\cite{bell:81}. Using a standard parabolic dispersion,
\be
\mu = E_{{\rm F}} = \frac{\hbar^2 k_{{\rm F}}^2}{2m^*}.
\ee
where $k_f$ is the Fermi wavevector and $m^*$ is the effective electron mass. For a 2D system, the number of states per unit area $N$  = $\frac{k_{{\rm F}}^2}{2\pi}$. Thus the DOS is well known as
\be
D(E_{{\rm F}}) = \frac{dN}{dE_{{\rm F}}} = \frac{m^*}{\pi\hbar^2}
\ee
From this relationship, we can infer that
\be
\frac{d\mu}{dD(E_{{\rm F}})} = \frac{dk_{{\rm F}}^2/2m^*}{dm^*} = -\frac{\pi \hbar^4 k_{{\rm F}}^2}{2m^{*2}}.
\ee
This result provides an important clue about the origin of a negative contribution to the compressibility: negative sign appears as chemical potential {\em decreases} with the effective mass per same carrier density. Therefore, to gain an understanding of when the electronic compressibility for a two-band system becomes negative, we need to understand the free energy of electron gas in the system.

Chemical potential is defined as the change in free energy over the change in the number of electrons $N_e$,
\be
\mu = \frac{dF}{dN} = \frac{df}{dn},
\ee
where $f$ = $F/A$ is the free energy divided by the area. One can relate the chemical potential to the density by normalizing the system by the area $A$, which allows us to define the electronic compressibility in terms of the free energy and the carrier density since $d\mu/dn$ = $d^2(f)/dn^2$,
\be
\kappa = \Big(n^2\frac{d^2f}{dn^2}\Big)^{-1}.
\ee
Since most experimental measurements examine the $d\mu / dn$, we will use inverse compressibility $\kappa^{-1}$ to discuss the crossover between positive and negative compressibility. 

{It is important to note that, in many systems, the effective mass and Fermi wavevector are direction dependent. Here, we  assume a simplified isotropic model.} 

\section{2D Electron Gas Free Energy}

\begin{figure}
\includegraphics[width=3.0in]{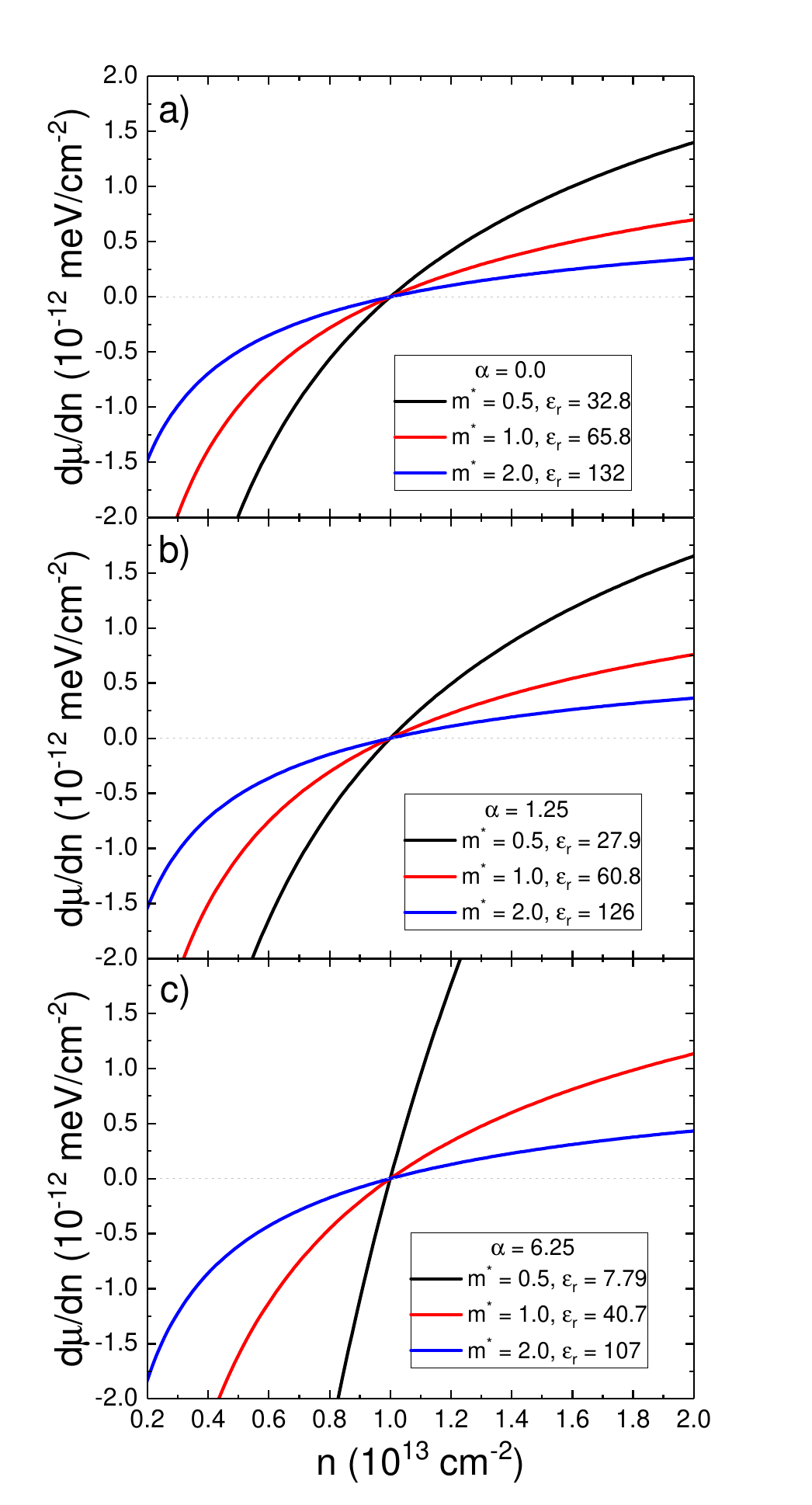}
\caption{(Color Online) {The inverse electronic compressibility for the one-band model plotted as a function of carrier density for various effective masses and out-of-plane electrostatic parameters $\alpha$ = 0 (a), 1.25 (b), and 6.25 (c). The quoted relative permittivities are given a the permittivity needed for the crossover carrier density to be $n_c$ = 1$\times$10$^{13}$ cm$^{-2}$.}}
\label{1band-dudn}
\end{figure}

Focusing on carrier density dependence of the free energy\cite{bell:81}, we decompose $f$ into kinetic\cite{kopp:09,tana:89}, exchange\cite{eise:94,rude:91}, and Coulomb components\cite{Scop:16,stef:17} ($f$ = $f_k$ + $f_{ex}$ + $f_c$) using an analytic model\cite{naga:84}, and examine each component individually for each band. Based on a homogeneous 2D electron model,
\be
\begin{array}{l}
\displaystyle f_{k} = \sum_i \frac{\pi n_i^2 \hbar^2}{2m_i^*}\\ \\
\displaystyle f_{ex} = -\sum_i \sqrt{\frac{2n_i^3}{\pi}}\frac{e^2}{3\pi\epsilon_{{\rm eff}}}\\\\
\displaystyle f_{c} = \sum_i n_i^2\frac{\alpha a_B e^2}{2\epsilon_{{\rm eff}}} - \sum_i \sqrt{\frac{n_i^3}{\pi}}\frac{e^2}{2\epsilon_{{\rm eff}}},\\
\end{array}
\ee
where the Coulomb or electrostatic term is broken into in-plane and out-of-plane components. Here, $\epsilon_{{\rm eff}}$ = $\epsilon_r \epsilon_0$ (the effective dielectric constant), and $e$ is the electron charge, $\alpha a_B$ is the effective distance between layers, $a_B$ is the Bohr radius 4$\pi \epsilon_0 \hbar^2/m^*e^2$ and $\alpha$ is a {phenomenological} tuning parameter for the out-of-plane electrostatic part of free energy. The out-of-plane electrostatic confinement term is needed for 2DEG materials with strong polarization. Overall, the full free energy is summed over $i$ bands, where $i=2$ is the specific case we consider. This free energy is a general analytic model and is meant to examine the general nature of the 2DEG in the absence of more complicated interactions. Therefore, we are using this model to detail qualitative features understand the differences between one- and two-band systems.

\begin{figure}[b]
\includegraphics[width=3.25in]{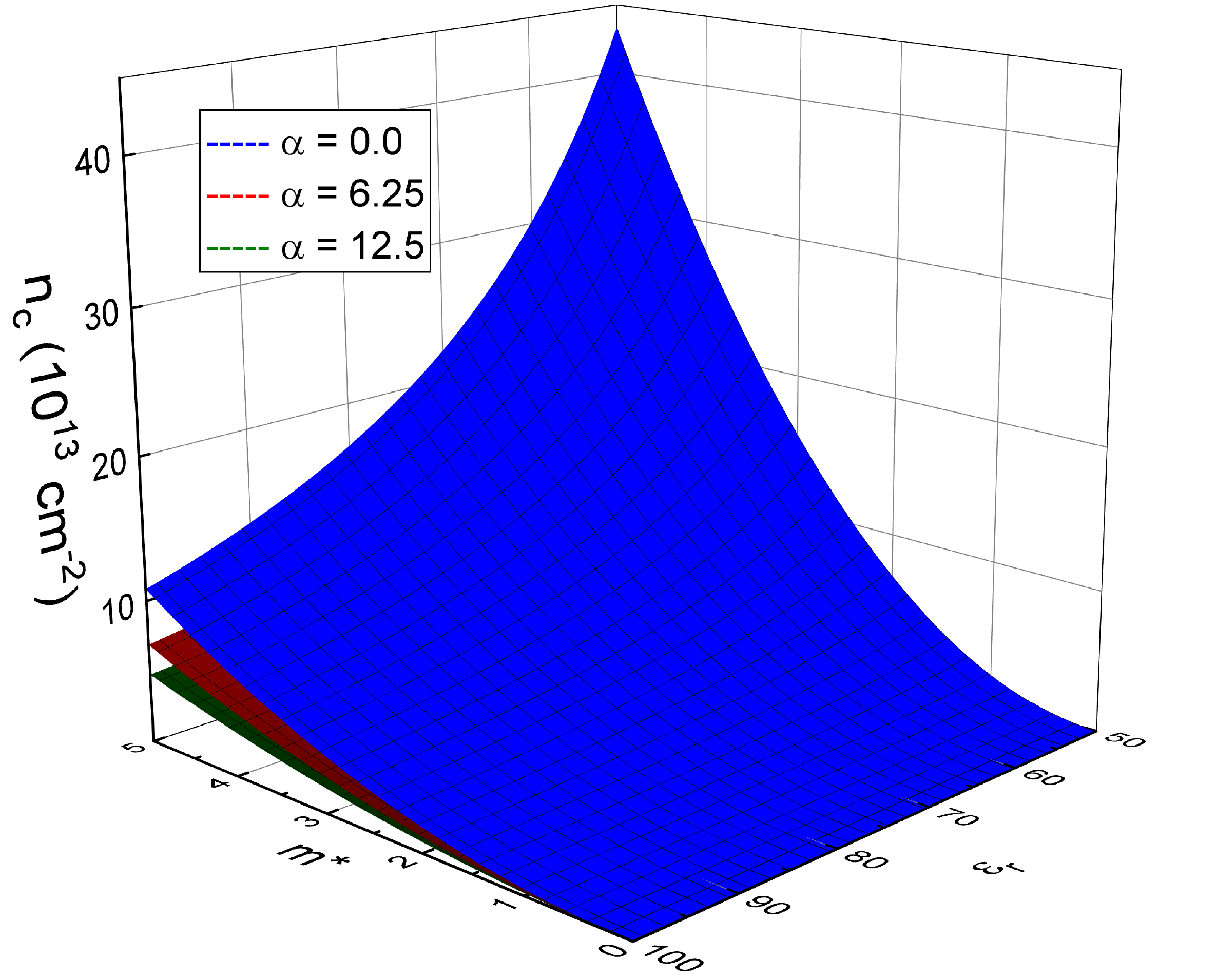}
\caption{(Color Online) {Crossover carrier density for the one-band model as a function of effective mass and relative permittivity for $\alpha$ = 0 (blue - top), 6.25 (red - middle), 12.5 (green - bottom).}}  
\label{1band-nc-d}
\end{figure}

\subsection{One-band model}

Using the free energy for a one-band 2DEG, the inverse electronic compressibility can be shown to be
\be
{\kappa^{-1} =n^2 \Big( {\frac {\pi{{\hbar}}^{2}}{{m^*}}}+{\frac {4\alpha{\pi}{{\hbar}}
^{2}}{{m^*}\,{\epsilon_r}}}-{\frac {3{e}^{2}}{8\sqrt {n\pi}
\epsilon_{eff}}}-\sqrt {\frac{2}{n\pi^{3}}}\frac{{e}^{2}}{4\epsilon_{eff}}\Big),}
\ee
where the effects of the parameters are shown in Fig. \ref{1band-dudn}. {Here, we present the inverse electronic compressibility as a function of the carrier density for various effective electron masses and relative permittivities as well as with and without the presence of out-of-plane electrostatic confinement. Here, we have chosen parameters for the effective mass and relative permittivity that will provide a crossover carrier density of 1.0 x 10$^{13}$ cm$^{-2}$ for various values of out-of-plane electrostatic parameter $\alpha$. The crossover points indicate carrier densities where electronic compressibility switches from positive and negative. These densities correspond to regime where which provides the general  a second band is populated. When compared to experimental results, it is not just the value of crossover carrier density value but also the line's overall slope through the point that also matters. The slope of the electronic compressibility through the critical point appears to be greatest when effective masses and permittivities are lowest, which indicates a greater increase in chemical potential with carrier density.} 

\begin{figure*}
\includegraphics[width=6.5in]{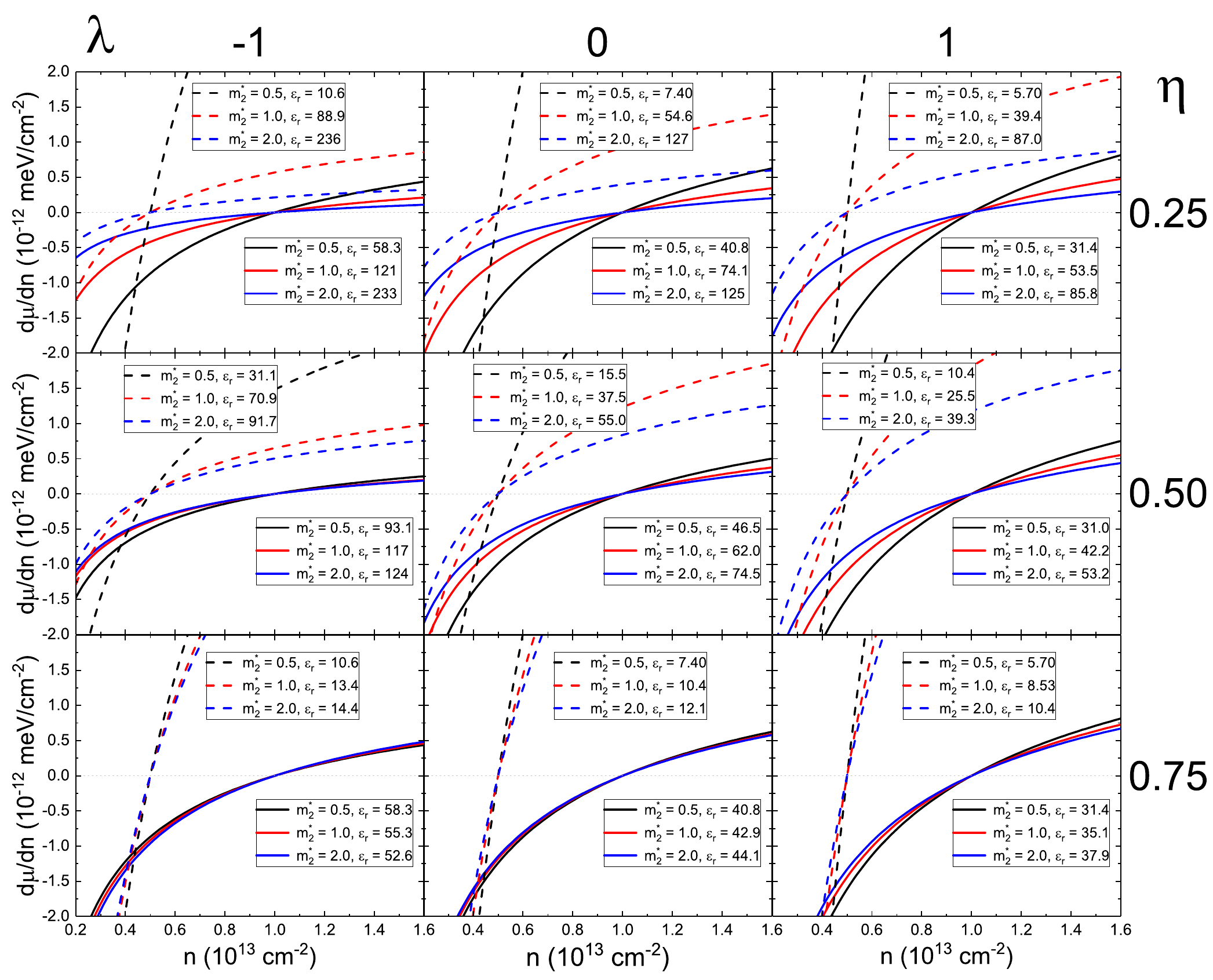}
\caption{(Color Online) {The inverse electronic compressibility as a function of carrier density for the 2-band system with various coupling interactions $\lambda$ = -1, 0 and 1 and density ratios $\eta$ = 0.25, 0.50, and 0.75. Within each plot, the inverse electronic compressibility is plotted for different effective masses. The out-of-plane electrostatic factor and the relative permittivity has been selected to make $n_c$ = 1.0 $\times$10$^{13}$ cm$^{-2}$ ($\alpha$ = 0.0) or 0.5$\times$10$^{13}$ cm$^{-2}$ ($\alpha$ = 12.5).}}
\label{2band-dudn}
\end{figure*}

Since the inverse electronic compressibility is proportional to the change in the chemical potential with regards to $n$, we can determine the critical crossover carrier density $n_c$ for the one-band model by solving $d\mu/dn$ = 0, and show that
\be
{n_c = \frac{(3\pi +2\sqrt{2})^2}{64\pi^{3}(\epsilon_r+4\alpha)^2}\Big(\frac{m^{*}e^2}{\pi\epsilon_{0}\hbar^2}\Big)^2}
\ee

{Here, the crossover carrier concentrations depend on the square of the effective mass over relative permittivity for the 2DEG. The addition of an out-of-plane electrostatic confinement field lowers the carrier density for negative electronic compressibility (shown in Fig. \ref{1band-nc-d}), where it is demonstrated that the crossover carrier density is controlled by effective mass, relative permittivity, and $z$-axis electrostatic potential. Within the general understanding of electrodynamics, an increasing effective mass will increase the number of states at the Fermi level, lowers the overall kinetic energy of the electronic system, and increases the critical carrier density for the transition.}

Figure \ref{1band-nc-d} shows that an increase in the relative permittivity leads to the decrease the critical carrier density. {The plots for larger values of $\alpha$ follow the same trend as the $\alpha$ = 0 curve with an overall lowering of the critical carrier density.} We interpret this effect being due to a decrease in the exchange and Coulomb interactions. This effect is also accompanied by increase of the out-of-plane electrostatic confinement.



These changes in the permittivity with carrier density are consistent with the general understanding of electronic structure. The lower the carrier density, the more insulating the material becomes, resulting in enhanced polarizability for the material. It is also shown that increasing the effective mass will also increase the needed permittivity due to the reduction in the conductivity, which will localize charge and allow for higher polarizabilities.


\begin{figure*}
\includegraphics[width=6.0in]{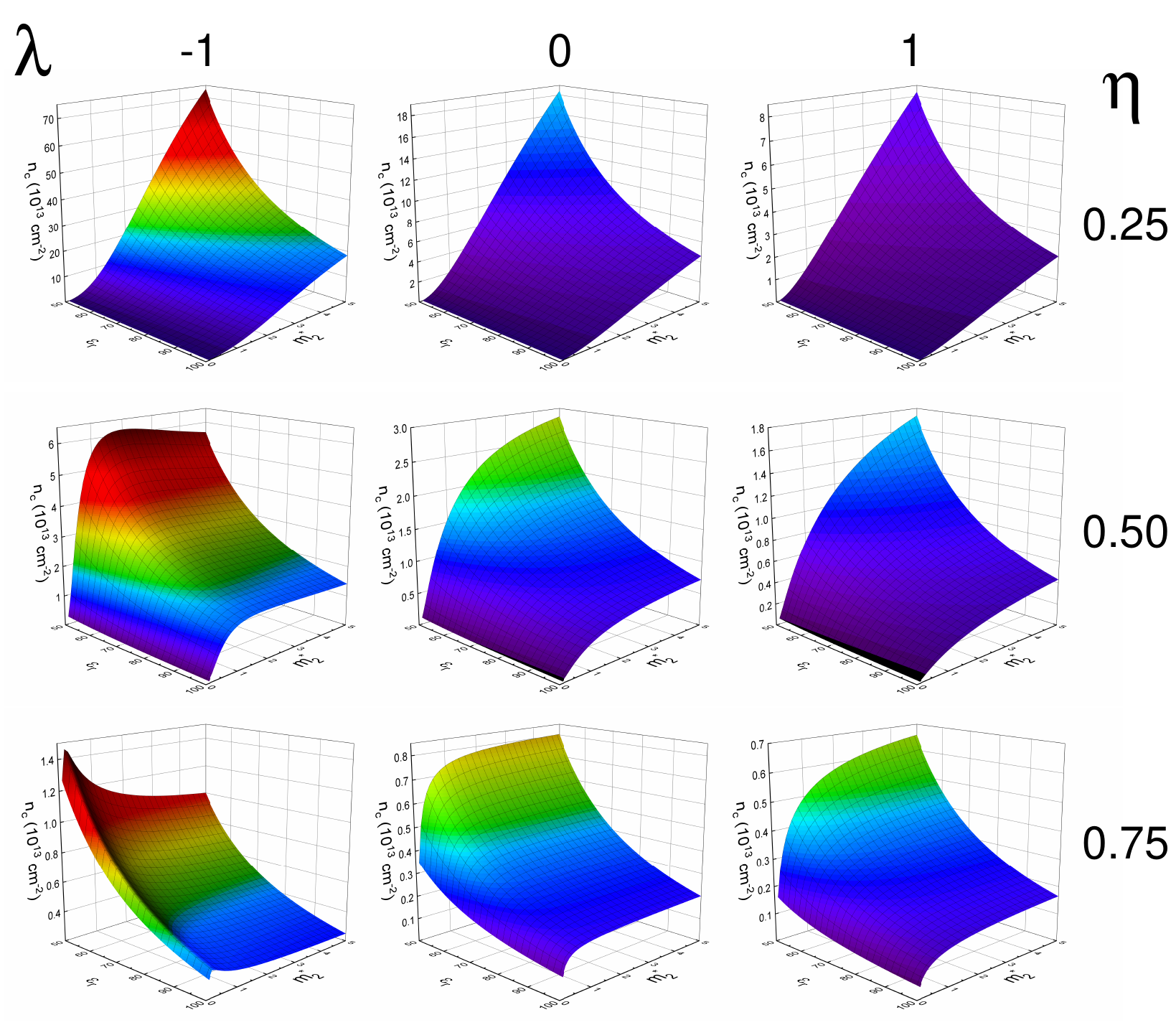}
\caption{(Color Online) {The critical carrier density as a function of $\epsilon_r$ and $m_2^*$ for different values of $\lambda$ and $\eta$ with $m_1^*$ = 0.5 and $\alpha$ = 0. The color scales for each row of plots are set by the $\lambda$ = -1 column to detail the the shift in carrier density.}}
\label{2band-nc-er-m}
\end{figure*}

\subsection{Two-band Model}

Since the crossover carrier density of 2DEG is a measure of band population, there should be distinct differences in the effects of a one-band system regarding a two-band system. Therefore, we generalize the previous analysis to a two-band case and slightly alter the model by adding the first-order coupling between bands. 

Assuming a two-band model ($i$ = 1,2), we have:
\be
\begin{array}{c}
\displaystyle f = \Big(\frac{\pi \hbar^2}{2} + \frac{{2 \pi \alpha \hbar^2 }}{\epsilon_r}\Big) \Big(\frac{n_1^2}{m_1^*} + \frac{n_2^2}{m_2^*}\Big)  \\\\
-\sqrt{\frac{2}{\pi^3}}\frac{e^2}{\epsilon_{{\rm eff}}}\left(\frac{1}{3}+\frac{\pi}{\sqrt{2^{3}}}\right)
\left(n_1^{\frac{3}{2}} + n_2^{\frac{3}{2}}\right) \displaystyle +  \frac{\lambda \pi \hbar^2}{2\sqrt{m_1^*m_2^*}} n_1n_2,
\end{array}
\ee
where $\lambda$ provides a first-order coupling between the two bands within a Ginzburg-Landau approximation, where this term is considered isotropic to help reduce the overall number of parameters. Since the interaction can be either repulsive or attractive, $\lambda$ can be either positive or negative. Here, the total interaction energy between two bands can be expanded as an ordered series of terms. We are considering only the first-order term, which is linear in both $n_1$ and $n_2$. Higher-order terms and other interactions could also be included, but they are typically system-specific and move us away from this more general model.

{From the free energy model, the inverse electronic compressibility} $\kappa^{-1}$ can now be written as
\be
\begin{array}{c}
\displaystyle \kappa^{-1} = {n}^{2} \Big[ \big( \pi \hbar^2 + \frac{{4 \pi} \alpha \hbar^2}{\epsilon_r}\big)\big(\frac{\eta^2}{m^*_1}+\frac{(1-\eta)^2}{m^*_2}\big) \\
- \frac{e^2}{4\epsilon_{{\rm eff}}\sqrt{n}}\big(\sqrt{\frac{2}{\pi^3}}+{\frac{3}{2\sqrt{\pi}}}\big)\big(\eta^{\frac{3}{2}}+(1-\eta)^{\frac{3}{2}}\big) + \frac{\lambda \pi \hbar^2 \big(\eta - \eta^2\big)}{\sqrt{m^*_1 m^*_2}}
\Big].
\end{array}
\ee
Here, we define the total carrier density $n$ = $n_1$ + $n_2$ and relative population factor $\eta$, {where $n_1$ = $\eta n$ and $n_2$ = $(1-\eta) n$}. Therefore, $\eta$ provides a simple gauge that changes the electron population from one band to the other {and is assumed to be independent on the carrier density.} In case of  $\eta$ = 1 or 0 the carrier population is solely in the first or second band. Any fractional concentration ratio describes carriers shared between bands;  $\eta$ = 0.5 corresponds to both bands being equally populated.

\begin{figure*}
\includegraphics[width=6.0in]{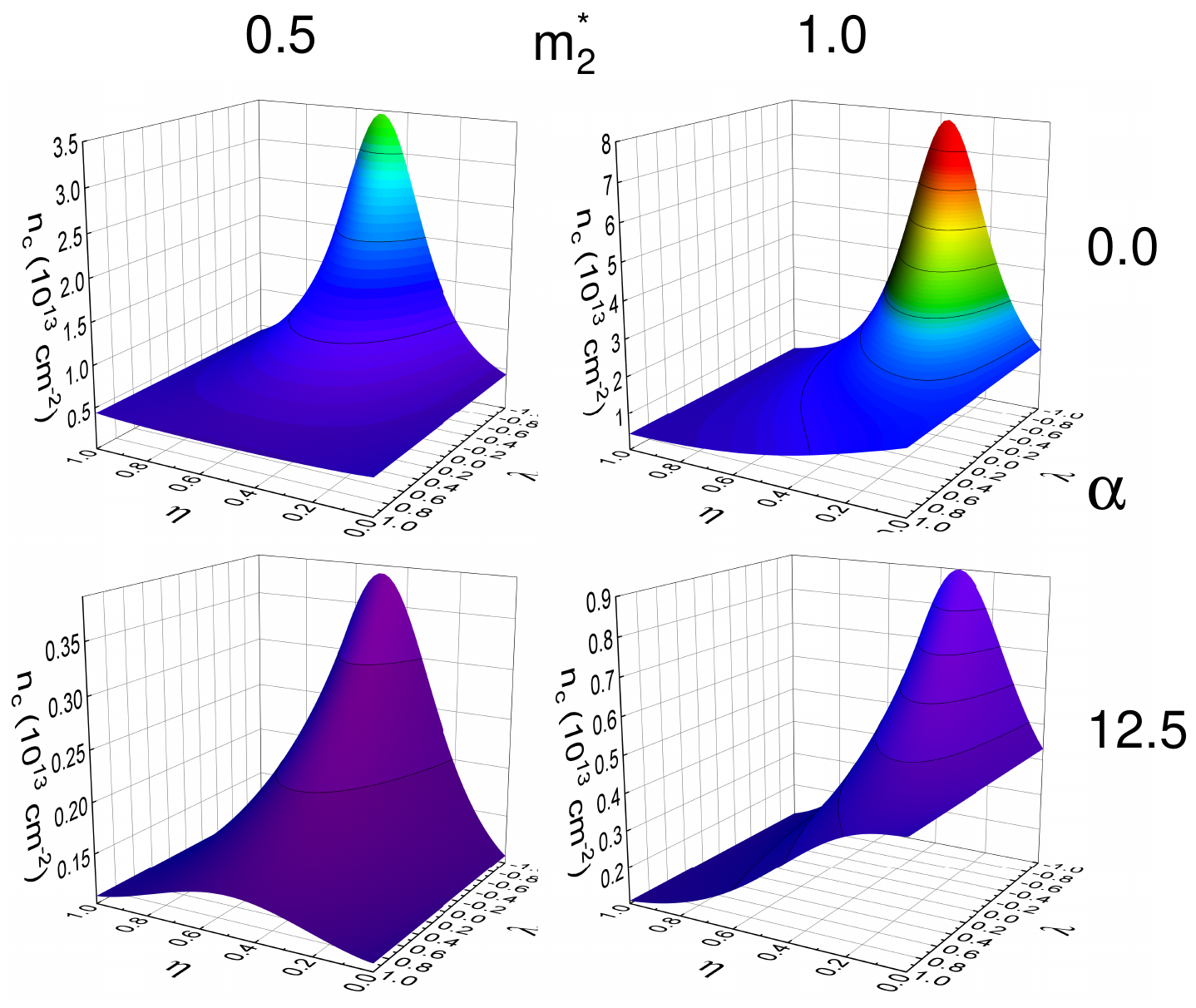}
\caption{(Color Online) {The effects of $\eta$ and $\lambda$ on the crossover carrier density with $m_2^*$ = 0.5 and 1.0 and $\alpha$ = 0.0 and 12.5. Here, $m_1^*$ is held at 0.5$m_e$ and $\epsilon_r$ = 50.}}
\label{2band-nc-d}
\end{figure*}

{As shown for the one-band model, Fig. \ref{2band-dudn} shows the inverse electronic compressibility as a function of the 2D carrier density for the two-band system with various coupling interactions $\lambda$, density ratios $\eta$, and effective masses. To illustrate the effect of $\alpha$, the data with critical crossover points at 1.0x10$^{13}$ cm$^{-2}$ have a value of $\alpha$ = 0, while 0.5x10$^{13}$ cm$^{-2}$ have a value of $\alpha$ = 12.5.}

{Similar to the one-band model, the lower effective masses produce a sharper slope of the curve through the critical points. The addition of a second band enhances the slope further as $\lambda$ or $\alpha$ are increased, where the dramatic increase in slope requires a lower effective dielectric constant. A general effect of the two-band model the need for lower overall polarizability to achieve the same crossover carrier density, which} leads to the suggestion that the combination of effective mass and dielectric measurements can help distinguish between one- or two-band systems.

\begin{figure*}
\includegraphics[width=6.0in]{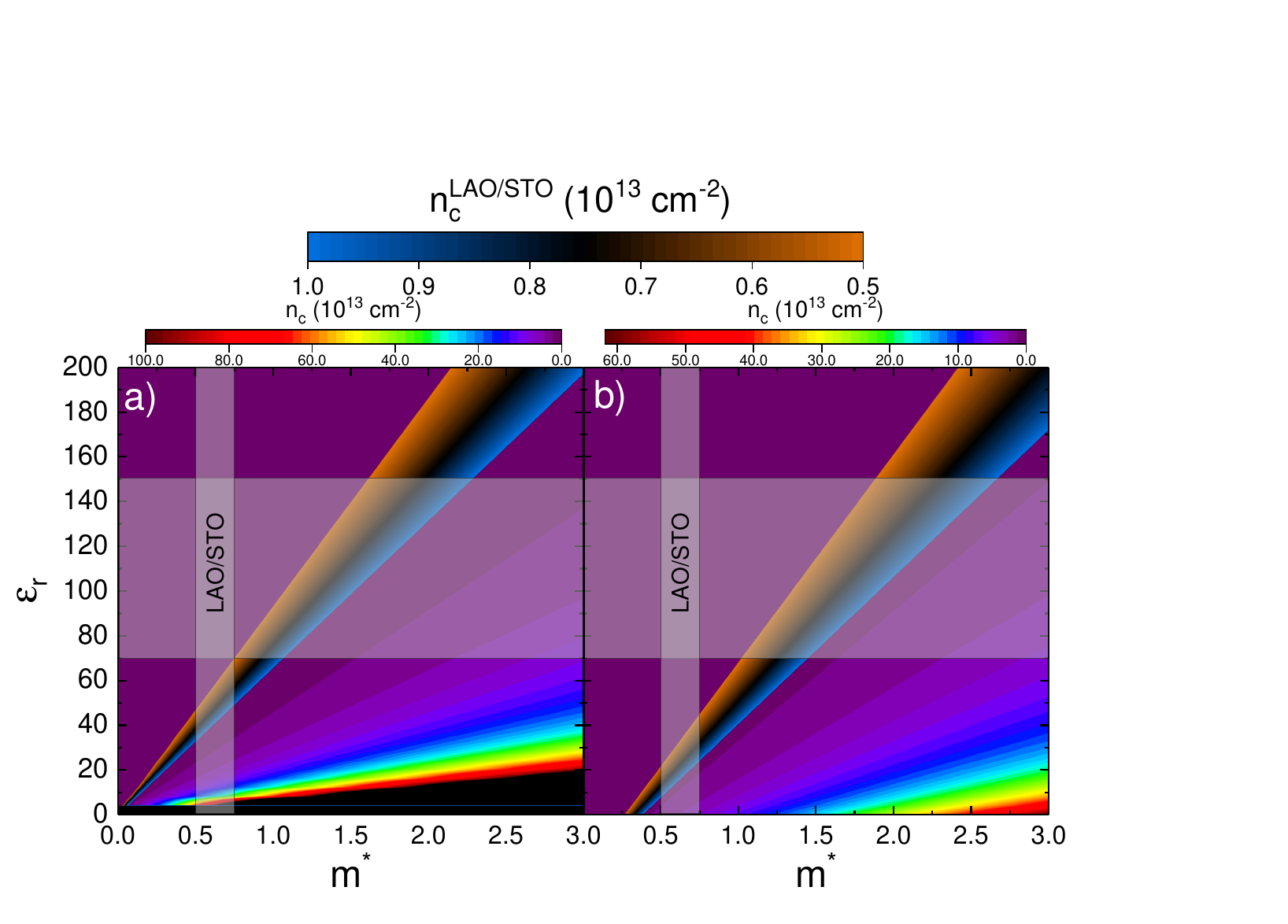}
\caption{(Color Online) {The critical cross-over carrier density is plotted as a function of relative permittivity and effective mass for the one-band model with $\alpha$ = 0 (a) and 12.5 (b). The shaded region illustrates the regime where most measurements of the relative permittivity and effective mass of LAO/STO are performed.  Within our model we conclude that for the known LAO/STO parameters does not fall within the predicted behavior of a one-band model.}}
\label{LAO-STO-range-1}
\end{figure*}

{To examine these effects  further, we focus on the crossover carrier density for the inverse electronic compressibility, given by}

{
\be
\begin{array}{ll}
 n_c =    &  \frac{(3\pi+2\sqrt{2})^2}{64\pi^{3}(\epsilon_r+4\alpha)^2}\Big(\frac{\textcolor{black}{\sqrt{m^{*}_1m^{*}_2}}e^2}{\pi\epsilon_{0}\hbar^2}\Big)^2\\
 &  \Big(\frac{\eta^{3/2}+(1-\eta)^{3/2}}{1/\gamma\eta^2 +\sqrt{\gamma}(1-\eta^2)+\lambda\eta(1-\eta)}\Big)^2
\label{2band_nc}
\end{array}
\ee}

{where $\gamma$ = $m_1^*$/$m_2^*$. Here, it is clear that the first part of the equation is the one-band model, and the second part details the second-band adjustment.} In a similar manner as the one-band model, Fig. \ref{2band-nc-er-m} shows the critical carrier density as a function of relative permittivity and effective mass of the second band with different values of $\lambda$ and $\eta$. Here, the first band's effective mass is set to 0.5 and $\alpha$ = 0. Similar trends in the one-band system also can be observed. Larger critical carrier density occurs in systems with larger effective masses and lowers relative permittivities.   Fig. \ref{2band-nc-er-m} illustrates the effect of  $\lambda$ and $\eta$ on the critical carrier density as a function of $\epsilon_r$ and $m_2^*$. 

In Fig. \ref{2band-nc-d}, the effects of $\eta$ and $\lambda$ on the crossover carrier density with $m_2^*$ = 0.5 and 1.0 and $\alpha$ = 0.0 and 12.5 are detailed. Here, $m_1^*$ is held at 0.5 and $\epsilon_r$ = 50, and shows that the sharing of carrier density between the bands increases the critical carrier density. This effect is illustrated best when the effective masses are equal. Here, the critical carrier density is symmetric about $\eta$ = 0.5, and $\eta$ results in shifting of electrons towards one band regime and produces higher carrier density regions with increasing effective mass. {It is important to note that the critical carrier density continues to increase beyond the $\lambda$ = -1 plane. Therefore, with respect to a given $\lambda$, the critical carrier density will peak at a given $\eta$, but for a given $\eta$, there is no global maximum.}

\begin{figure*}
\includegraphics[width=6.0in]{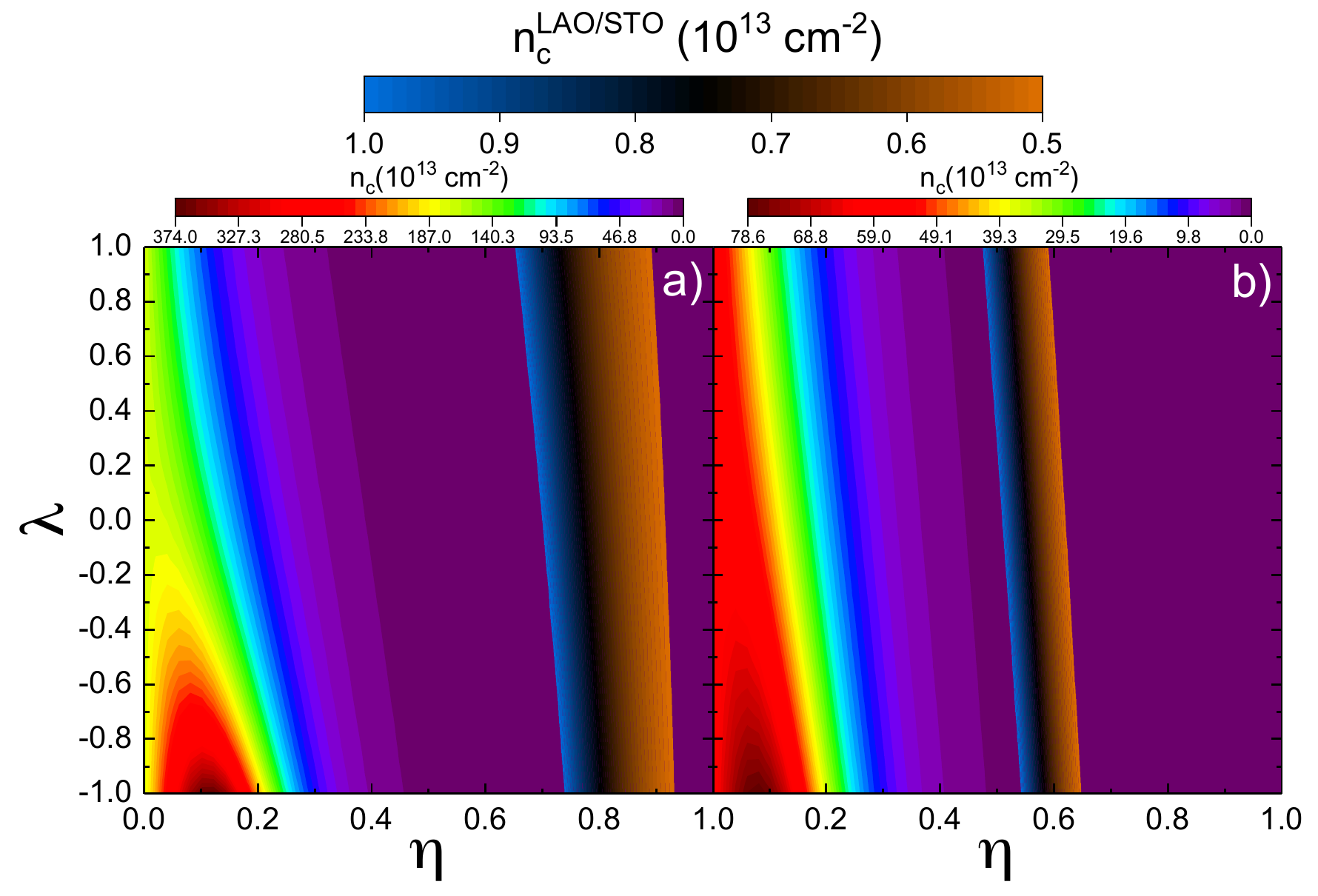}
\caption{(Color Online) {The critical cross-over carrier density is plotted as a function of $\lambda$ and $\eta$ for the closest LAO/STO parameters ($\epsilon_r$ = 70, $m_1^*$ = 0.7, and $m_2^*$ = 14) for the two-band model with $\alpha$ = 0 (a) and 12.5 (b). The blue/brown cone region denotes the LAO/STO critical carrier density region and provides a multi-band parameter space for experimentalists to investigate the presence of the two-band electronic states in these devices.}}
\label{LAO-STO-range-2}
\end{figure*}

From Eq. \ref{2band_nc}, we see some overall trends:
i) as the density ratio is shifted from the equal point of {$\eta$ = 0.5, the electronic system requires} an overall increase permittivity and becomes more apparent with a negative $\lambda$. ii) as $\lambda$ increases in value, the {needed} relative permittivity is decreased. While the overall dependence of the critical carrier density from the effective permittivity and masses is similar to the one-band model, examining the combination of measured parameters can help determine the likelihood of either a one- or two-band system.

By looking at the {$\eta$ = 0.5 and $\lambda$ = 0} case in Fig. \ref{2band-dudn}, the critical carrier density response is quite similar to that of the one-band model. If one includes band interactions the crossover points shift depending on the type of interaction and strength: for increasing $\lambda$, the critical point {shifts as the required permittivity decreases}.


To futher examine the effects of multi-band interactions, we focus on the example of the LAO/STO interface, {where some experiments have indicated that the system is possibly in the multi-band regime\cite{binn:80,josh:12,sant:11}.}

\section{Relevance to LAO/STO interfaces}

The interface of LaAlO$_3$ (LAO) and SrTiO$_3$ (STO) has become a textbook example of the emergence of complex phenomena at complex oxides' interfaces where two band insulators produce a 2DEG, which becomes superconducting at 200-300 mK\cite{ohto:04,reyr:07,bisc:12}. Studies into the nature of the superconducting state at the LAO/STO interface has revealed the potential for a multi-band electronic system\cite{fern:13,smin:17,veer:13,gudu:13,vanm:08}. On the other hand, there is a significant body of literature that seems to indicate that the interfacial electronic states correspond to a single band \cite{huij:09, Gari:09, vand:11,gari:15}. {Therefore, to provide a method that could help identify mutli-band electronic systems, we present an approach based on observable properties of the normal electronic states. 
However, the difference occurs in the material parameters' values, which is one reason multi-band systems are challenging to identify. It is not in the curvature or functional form of the potential, but a precise analysis of measured parameters.}

Electron-doped n-type STO exhibits a distinct transition into a superconductor with a peak $T_c$ = 200 mK. Moreover, theoretical and experimental studies have shown that the superconducting transition can be dramatically enhanced under pressure\cite{dunn:18,herr:19}.  In Fernandes et al.\cite{fern:13} we described multi-band aspects of this superconductivity and discussed the similarities between bulk STO and the LAO/STO interface. In the present work we adopt this two-band model for general case of normal state 2DEGs to LAO/STO interface  to determine the most relevant features. 

To make a relevant comparison  we would need to determine values of dielectric constant for interfaces. For the bulk systems, LAO has a measured dielectric constant of about 18-24\cite{bark:11,nevi:72}. In contrast, STO can have a dielectric constant upwards of 25000 at zero electric fields, and drops to around 300 (with an electric field or at high temperatures)\cite{bark:11,berg:95}. Since the 2DEG exists in the STO layer with a typical depth of 5-7 nm\cite{sing:09}, the results of the analysis of the two-band model depend greatly on the measured value of the dielectric constant and the effective mass. To be specific, we assume a range of dielectric constants between about 70 and 150\cite{jano:11,peel:15,copi:09,mazn:18,lee:18,Guo:16,zhon:13,mcco:14,rein:12,song:18} and an estimated effective masses for the LAO/STO range from 0.5$m_e$-0.7$m_e$ for the first band and 5.0-14$m_e$ for the second band\cite{jano:11}.

In LAO/STO, the critical carrier density is very device-dependent and seems to fall between 0.5 and 1.0 x 10$^{13}$ cm$^{-2}$.\cite{li:11}. With these ranges in mind, Fig. \ref{LAO-STO-range-1} shows the carrier density as a function of relative permittivity and effective mass for two different electrostatic parameters for the one-band model. The parameter ranges for LAO/STO have been shaded in grey. Through the middle, the brown contour shows the permittivity and effective mass range for the general critical carrier density range of LAO/STO. In Fig. \ref{LAO-STO-range-1}(a), the measured LAO/STO parameters come very close to falling into the carrier density range for $\alpha$ = 0. However, suppose one includes an out-of-plane electrostatic energy (Fig. \ref{LAO-STO-range-1}(b)), which is needed for the polar heterostructure of LAO/STO. In that case, the parameters become even further apart. Therefore, given the wide range of device-specific measurements, it seems that the one-band model is not applicable.

Figure \ref{LAO-STO-range-2} shows the carrier density as a function of relative permittivity and effective mass for two different electrostatic parameters for the two-band model. The main issue is that we do not know the relative population of the bands nor the coupling constant. Therefore, by using the closest parameters from the one-band model ($\epsilon_r$ = 70 and $m_1^*$ = 0.7$m_e$) and instituting a second-band effective mass of 14$m_e$, we show the range of $\eta$ and $\lambda$ needed to provide critical carrier density of between 0.5 and 1.0 x 10$^{13}$ cm$^{-2}$ for $\alpha$ = 0.0 and 12.5 (without and with electrostatic confinement).

We conclude that the band filling never falls on $\eta$ = 1 or 0, hence we assume that a two-band model is needed to account for the measured electronic parameters of LAO/STO. Furthermore, if the system demonstrates larger effective mass or a lower permittivity, then the LAO/STO interface may reduce back to a one-band model (as shown in Fig. \ref{LAO-STO-range-1}), which \textcolor{black}{is} why we have chosen parameters that appear to be on the limits of measured quantities for this system. \textcolor{black}{Our model allows the direct comparison of physical parameters for single vs multi-band systems and thus provides a useful tool for experimentalists to investigate the multi-band systems.}

Our approach demonstrates that effective masses for the 2DEG must be reasonably small for the first band to produce a similar compressibility slope and dielectric constant. We find more flexibility with the inter-band coupling and variable carrier density within a two-band model.

Overall, the proposed  two-band 2DEG model demonstrates that stronger inter-band coupling lowers the required $\epsilon_r$ for the system. The coupling of the bands produces less electron screening. It makes the system less metallic, consistent with Fernandes $et~al.$\cite{fern:13}, where it was shown that a weakly interacting two-band model adequately describes the 2D STO system. {Our our analys indicates that, within the 2DEG model, a two-band or multi-band system is adequately describing the LAO/STO interface \cite{binn:80,fern:13}. We also mention multiple experimental observations suggesting multiband natire of 2DEG \cite{song:18,smin:17,veer:13,gudu:13}.}

{It should be mentioned that this analysis is done with an isotropic model. However, in the LAO/STO system, the second band being occupied a doubly degenerate band with a fairly anisotropic effective mass that can range from 5 to 14 $m_e$ depending on the orientation\cite{josh:12,smin:18}. Therefore, in our analysis, we have chosen to use a large effective mass of 14$m_e$ for the second band, which illustrates that the system still prefers the multi-band model over the single band.}

\section{Conclusions}

In this paper, {we provide a general method for determining the effects of multi-band interactions on a two-dimensional electron gas through  electronic compressibility.} Using a 2D electron gas model with a first-order inter-band coupling between the carrier densities, we examine the dependence of negative compressibility of the 2D electron gas on critical carrier density, dielectric constant, and effective mass and compare the results to the interface of LaAlO$_3$/SrTiO$_3$. {Our calculations show that the presence of inter-band coupling has an effect on the polarizability and the critical carrier density of the 2DEG.} Furthermore, we find that the bands' effective masses have a distinct and dramatic effect on the negative compressibility.

Given recent suggestions that the LAO/STO interface is a multi-band electron system\cite{fern:13,smin:17,veer:13,gudu:13}, we relate  the negative compressibility to inter-band and intra-band interactions and to the effective electron masses in the bands. We find that a one-band model does not  reproduce the LAO/STO interface's critical carrier density. On the other hand, using a  two-band model, we find the critical carrier density for the negative compressibility crossover of LAO/STO within the parameter range that  consistent with the observations.

{Among future applications of the model presented here, we mention the effects of dynamics on electronic properties like transport and compressibility.  For example, the role of dynamics and pumping in producing transient negative compressibility regime in multi-band 2DEG would be an interesting problem to pursue.  We also point to the important questions about the nature of the superconducting state, seen in LAO/STO, in the presence of two coupled electronic bands.}

\section*{Acknowledgements}

{The author would like to thank the referees, especially the second referee, for their insight and helpful comments. Additionally,} A.M. and J.T.H. would like to acknowledge support by the Institute for Materials Science at Los Alamos National Laboratory. The Work of A.V.B. was supported by the University of Connecticut, VILLUM FONDEN via the Centre of Excellence for Dirac Materials (Grant No. 11744), the European Research Council under the European Unions Seventh Framework Program Synergy HERO, and KAW.

\bibliography{LAOSTO}

\providecommand{\noopsort}[1]{}\providecommand{\singleletter}[1]{#1}%
\begin{thebibliography}{76}%
\makeatletter
\providecommand \@ifxundefined [1]{%
 \@ifx{#1\undefined}
}%
\providecommand \@ifnum [1]{%
 \ifnum #1\expandafter \@firstoftwo
 \else \expandafter \@secondoftwo
 \fi
}%
\providecommand \@ifx [1]{%
 \ifx #1\expandafter \@firstoftwo
 \else \expandafter \@secondoftwo
 \fi
}%
\providecommand \natexlab [1]{#1}%
\providecommand \enquote  [1]{``#1''}%
\providecommand \bibnamefont  [1]{#1}%
\providecommand \bibfnamefont [1]{#1}%
\providecommand \citenamefont [1]{#1}%
\providecommand \href@noop [0]{\@secondoftwo}%
\providecommand \href [0]{\begingroup \@sanitize@url \@href}%
\providecommand \@href[1]{\@@startlink{#1}\@@href}%
\providecommand \@@href[1]{\endgroup#1\@@endlink}%
\providecommand \@sanitize@url [0]{\catcode `\\12\catcode `\$12\catcode
  `\&12\catcode `\#12\catcode `\^12\catcode `\_12\catcode `\%12\relax}%
\providecommand \@@startlink[1]{}%
\providecommand \@@endlink[0]{}%
\providecommand \url  [0]{\begingroup\@sanitize@url \@url }%
\providecommand \@url [1]{\endgroup\@href {#1}{\urlprefix }}%
\providecommand \urlprefix  [0]{URL }%
\providecommand \Eprint [0]{\href }%
\providecommand \doibase [0]{https://doi.org/}%
\providecommand \selectlanguage [0]{\@gobble}%
\providecommand \bibinfo  [0]{\@secondoftwo}%
\providecommand \bibfield  [0]{\@secondoftwo}%
\providecommand \translation [1]{[#1]}%
\providecommand \BibitemOpen [0]{}%
\providecommand \bibitemStop [0]{}%
\providecommand \bibitemNoStop [0]{.\EOS\space}%
\providecommand \EOS [0]{\spacefactor3000\relax}%
\providecommand \BibitemShut  [1]{\csname bibitem#1\endcsname}%
\let\auto@bib@innerbib\@empty
\bibitem [{\citenamefont {Heber}(2009)}]{hebe:09}%
  \BibitemOpen
  \bibfield  {author} {\bibinfo {author} {\bibfnamefont {J.}~\bibnamefont
  {Heber}},\ }\bibfield  {title} {\bibinfo {title} {Materials science: Enter
  the oxides},\ }\href {https://doi.org/10.1038/459028a} {\bibfield  {journal}
  {\bibinfo  {journal} {Nature}\ }\textbf {\bibinfo {volume} {459}},\ \bibinfo
  {pages} {28} (\bibinfo {year} {2009})}\BibitemShut {NoStop}%
\bibitem [{\citenamefont {Prellier}\ \emph {et~al.}(2005)\citenamefont
  {Prellier}, \citenamefont {Singh},\ and\ \citenamefont
  {Murugavel}}]{prel:05}%
  \BibitemOpen
  \bibfield  {author} {\bibinfo {author} {\bibfnamefont {W.}~\bibnamefont
  {Prellier}}, \bibinfo {author} {\bibfnamefont {M.}~\bibnamefont {Singh}},\
  and\ \bibinfo {author} {\bibfnamefont {P.}~\bibnamefont {Murugavel}},\
  }\bibfield  {title} {\bibinfo {title} {The single-phase multiferroic oxides:
  From bulk to thin film},\ }\href
  {https://doi.org/10.1088/0953-8984/17/30/R01} {\bibfield  {journal} {\bibinfo
   {journal} {Journal of Physics: Condensed Matter}\ }\textbf {\bibinfo
  {volume} {17}},\ \bibinfo {pages} {R803} (\bibinfo {year}
  {2005})}\BibitemShut {NoStop}%
\bibitem [{\citenamefont {Khomskii}\ and\ \citenamefont
  {Sawatzky}(1997)}]{khom:97}%
  \BibitemOpen
  \bibfield  {author} {\bibinfo {author} {\bibfnamefont {D.}~\bibnamefont
  {Khomskii}}\ and\ \bibinfo {author} {\bibfnamefont {G.}~\bibnamefont
  {Sawatzky}},\ }\bibfield  {title} {\bibinfo {title} {Interplay between spin,
  charge and orbital degrees of freedom in magnetic oxides},\ }\href
  {https://doi.org/10.1016/S0038-1098(96)00717-X} {\bibfield  {journal}
  {\bibinfo  {journal} {Solid State Communications}\ }\textbf {\bibinfo
  {volume} {102}},\ \bibinfo {pages} {87} (\bibinfo {year} {1997})}\BibitemShut
  {NoStop}%
\bibitem [{\citenamefont {Zubko}\ \emph {et~al.}(2011)\citenamefont {Zubko},
  \citenamefont {Gariglio}, \citenamefont {Gabay}, \citenamefont {Ghosez},\
  and\ \citenamefont {Triscone}}]{zubk:11}%
  \BibitemOpen
  \bibfield  {author} {\bibinfo {author} {\bibfnamefont {P.}~\bibnamefont
  {Zubko}}, \bibinfo {author} {\bibfnamefont {S.}~\bibnamefont {Gariglio}},
  \bibinfo {author} {\bibfnamefont {M.}~\bibnamefont {Gabay}}, \bibinfo
  {author} {\bibfnamefont {P.}~\bibnamefont {Ghosez}},\ and\ \bibinfo {author}
  {\bibfnamefont {J.-M.}\ \bibnamefont {Triscone}},\ }\bibfield  {title}
  {\bibinfo {title} {Interface physics in complex oxide heterostructures},\
  }\href {https://doi.org/10.1146/annurev-conmatphys-062910-140445} {\bibfield
  {journal} {\bibinfo  {journal} {Annu. Rev. Condens. Matter Phys.}\ }\textbf
  {\bibinfo {volume} {2}},\ \bibinfo {pages} {141} (\bibinfo {year}
  {2011})}\BibitemShut {NoStop}%
\bibitem [{\citenamefont {Chakhalian}\ \emph {et~al.}(2007)\citenamefont
  {Chakhalian}, \citenamefont {Freeland}, \citenamefont {Habermeier},
  \citenamefont {Cristiani}, \citenamefont {Khaliullin}, \citenamefont {van
  Veenendaal},\ and\ \citenamefont {Keimer}}]{chak:07}%
  \BibitemOpen
  \bibfield  {author} {\bibinfo {author} {\bibfnamefont {J.}~\bibnamefont
  {Chakhalian}}, \bibinfo {author} {\bibfnamefont {J.}~\bibnamefont
  {Freeland}}, \bibinfo {author} {\bibfnamefont {H.}~\bibnamefont
  {Habermeier}}, \bibinfo {author} {\bibfnamefont {G.}~\bibnamefont
  {Cristiani}}, \bibinfo {author} {\bibfnamefont {G.}~\bibnamefont
  {Khaliullin}}, \bibinfo {author} {\bibfnamefont {M.}~\bibnamefont {van
  Veenendaal}},\ and\ \bibinfo {author} {\bibfnamefont {B.}~\bibnamefont
  {Keimer}},\ }\bibfield  {title} {\bibinfo {title} {Orbital reconstruction and
  covalent bonding at an oxide interface},\ }\href
  {https://doi.org/10.1126/science.1149338} {\bibfield  {journal} {\bibinfo
  {journal} {Science (New York, N.Y.)}\ }\textbf {\bibinfo {volume} {318}},\
  \bibinfo {pages} {1114} (\bibinfo {year} {2007})}\BibitemShut {NoStop}%
\bibitem [{\citenamefont {Lee}\ \emph {et~al.}(2018)\citenamefont {Lee},
  \citenamefont {Campbell}, \citenamefont {Lee}, \citenamefont {Asel},
  \citenamefont {Paudel}, \citenamefont {Zhou}, \citenamefont {Lee},
  \citenamefont {Noesges}, \citenamefont {Seo}, \citenamefont {Park},
  \citenamefont {Brillson}, \citenamefont {Oh}, \citenamefont {Tsymbal},
  \citenamefont {Rzchowski},\ and\ \citenamefont {Eom}}]{lee:18}%
  \BibitemOpen
  \bibfield  {author} {\bibinfo {author} {\bibfnamefont {H.}~\bibnamefont
  {Lee}}, \bibinfo {author} {\bibfnamefont {N.}~\bibnamefont {Campbell}},
  \bibinfo {author} {\bibfnamefont {J.}~\bibnamefont {Lee}}, \bibinfo {author}
  {\bibfnamefont {T.}~\bibnamefont {Asel}}, \bibinfo {author} {\bibfnamefont
  {T.}~\bibnamefont {Paudel}}, \bibinfo {author} {\bibfnamefont
  {H.}~\bibnamefont {Zhou}}, \bibinfo {author} {\bibfnamefont {J.}~\bibnamefont
  {Lee}}, \bibinfo {author} {\bibfnamefont {B.}~\bibnamefont {Noesges}},
  \bibinfo {author} {\bibfnamefont {J.}~\bibnamefont {Seo}}, \bibinfo {author}
  {\bibfnamefont {B.}~\bibnamefont {Park}}, \bibinfo {author} {\bibfnamefont
  {L.}~\bibnamefont {Brillson}}, \bibinfo {author} {\bibfnamefont
  {S.}~\bibnamefont {Oh}}, \bibinfo {author} {\bibfnamefont {E.}~\bibnamefont
  {Tsymbal}}, \bibinfo {author} {\bibfnamefont {M.}~\bibnamefont {Rzchowski}},\
  and\ \bibinfo {author} {\bibfnamefont {C.-B.}\ \bibnamefont {Eom}},\
  }\bibfield  {title} {\bibinfo {title} {Direct observation of a
  two-dimensional hole gas at oxide interfaces},\ }\href
  {https://doi.org/10.1038/s41563-017-0002-4} {\bibfield  {journal} {\bibinfo
  {journal} {Nature Materials}\ }\textbf {\bibinfo {volume} {17}} (\bibinfo
  {year} {2018})}\BibitemShut {NoStop}%
\bibitem [{\citenamefont {Reinle-Schmitt}\ \emph {et~al.}(2012)\citenamefont
  {Reinle-Schmitt}, \citenamefont {Cancellieri}, \citenamefont {Li},
  \citenamefont {Fontaine}, \citenamefont {Medarde}, \citenamefont
  {Pomjakushina}, \citenamefont {Schneider}, \citenamefont {Gariglio},
  \citenamefont {Ghosez}, \citenamefont {Triscone},\ and\ \citenamefont
  {Willmott}}]{rein:12}%
  \BibitemOpen
  \bibfield  {author} {\bibinfo {author} {\bibfnamefont {M.}~\bibnamefont
  {Reinle-Schmitt}}, \bibinfo {author} {\bibfnamefont {C.}~\bibnamefont
  {Cancellieri}}, \bibinfo {author} {\bibfnamefont {D.}~\bibnamefont {Li}},
  \bibinfo {author} {\bibfnamefont {D.}~\bibnamefont {Fontaine}}, \bibinfo
  {author} {\bibfnamefont {M.}~\bibnamefont {Medarde}}, \bibinfo {author}
  {\bibfnamefont {E.}~\bibnamefont {Pomjakushina}}, \bibinfo {author}
  {\bibfnamefont {C.}~\bibnamefont {Schneider}}, \bibinfo {author}
  {\bibfnamefont {S.}~\bibnamefont {Gariglio}}, \bibinfo {author}
  {\bibfnamefont {P.}~\bibnamefont {Ghosez}}, \bibinfo {author} {\bibfnamefont
  {J.-M.}\ \bibnamefont {Triscone}},\ and\ \bibinfo {author} {\bibfnamefont
  {P.}~\bibnamefont {Willmott}},\ }\bibfield  {title} {\bibinfo {title}
  {Tunable conductivity threshold at polar oxide interfaces},\ }\href
  {https://doi.org/10.1038/ncomms1936} {\bibfield  {journal} {\bibinfo
  {journal} {Nature communications}\ }\textbf {\bibinfo {volume} {3}},\
  \bibinfo {pages} {932} (\bibinfo {year} {2012})}\BibitemShut {NoStop}%
\bibitem [{\citenamefont {Haraldsen}\ \emph {et~al.}(2012)\citenamefont
  {Haraldsen}, \citenamefont {W\"olfle},\ and\ \citenamefont
  {Balatsky}}]{hara:12}%
  \BibitemOpen
  \bibfield  {author} {\bibinfo {author} {\bibfnamefont {J.~T.}\ \bibnamefont
  {Haraldsen}}, \bibinfo {author} {\bibfnamefont {P.}~\bibnamefont
  {W\"olfle}},\ and\ \bibinfo {author} {\bibfnamefont {A.~V.}\ \bibnamefont
  {Balatsky}},\ }\bibfield  {title} {\bibinfo {title} {Understanding the
  electric-field enhancement of the superconducting transition temperature for
  complex oxide interfaces},\ }\href
  {https://doi.org/10.1103/PhysRevB.85.134501} {\bibfield  {journal} {\bibinfo
  {journal} {Phys. Rev. B}\ }\textbf {\bibinfo {volume} {85}},\ \bibinfo
  {pages} {134501} (\bibinfo {year} {2012})}\BibitemShut {NoStop}%
\bibitem [{\citenamefont {Kozuka}\ \emph {et~al.}(2011)\citenamefont {Kozuka},
  \citenamefont {Tsukazaki}, \citenamefont {Maryenko}, \citenamefont {Falson},
  \citenamefont {Akasaka}, \citenamefont {Nakahara}, \citenamefont {Nakamura},
  \citenamefont {Awaji}, \citenamefont {Ueno},\ and\ \citenamefont
  {Kawasaki}}]{kozu:11}%
  \BibitemOpen
  \bibfield  {author} {\bibinfo {author} {\bibfnamefont {Y.}~\bibnamefont
  {Kozuka}}, \bibinfo {author} {\bibfnamefont {A.}~\bibnamefont {Tsukazaki}},
  \bibinfo {author} {\bibfnamefont {D.}~\bibnamefont {Maryenko}}, \bibinfo
  {author} {\bibfnamefont {J.}~\bibnamefont {Falson}}, \bibinfo {author}
  {\bibfnamefont {S.}~\bibnamefont {Akasaka}}, \bibinfo {author} {\bibfnamefont
  {K.}~\bibnamefont {Nakahara}}, \bibinfo {author} {\bibfnamefont
  {S.}~\bibnamefont {Nakamura}}, \bibinfo {author} {\bibfnamefont
  {S.}~\bibnamefont {Awaji}}, \bibinfo {author} {\bibfnamefont
  {K.}~\bibnamefont {Ueno}},\ and\ \bibinfo {author} {\bibfnamefont
  {M.}~\bibnamefont {Kawasaki}},\ }\bibfield  {title} {\bibinfo {title}
  {{Insulating phase of a two-dimensional electron gas in
  Mg$_{x}$Zn$_{1-x}$O/ZnO heterostructures below $\nu = \frac{1}{3}$}},\ }\href
  {https://doi.org/10.1103/PhysRevB.84.033304} {\bibfield  {journal} {\bibinfo
  {journal} {Phys. Rev. B}\ }\textbf {\bibinfo {volume} {84}},\ \bibinfo
  {pages} {033304} (\bibinfo {year} {2011})}\BibitemShut {NoStop}%
\bibitem [{\citenamefont {Park}\ \emph {et~al.}(2010)\citenamefont {Park},
  \citenamefont {Bogorin}, \citenamefont {Cen}, \citenamefont {Felker},
  \citenamefont {Zhang}, \citenamefont {Nelson}, \citenamefont {Bark},
  \citenamefont {Folkman}, \citenamefont {Pan}, \citenamefont {Rzchowski},
  \citenamefont {Levy},\ and\ \citenamefont {Eom}}]{park:10}%
  \BibitemOpen
  \bibfield  {author} {\bibinfo {author} {\bibfnamefont {J.~W.}\ \bibnamefont
  {Park}}, \bibinfo {author} {\bibfnamefont {D.}~\bibnamefont {Bogorin}},
  \bibinfo {author} {\bibfnamefont {C.}~\bibnamefont {Cen}}, \bibinfo {author}
  {\bibfnamefont {D.}~\bibnamefont {Felker}}, \bibinfo {author} {\bibfnamefont
  {Y.}~\bibnamefont {Zhang}}, \bibinfo {author} {\bibfnamefont
  {C.}~\bibnamefont {Nelson}}, \bibinfo {author} {\bibfnamefont {C.~W.}\
  \bibnamefont {Bark}}, \bibinfo {author} {\bibfnamefont {C.}~\bibnamefont
  {Folkman}}, \bibinfo {author} {\bibfnamefont {X.}~\bibnamefont {Pan}},
  \bibinfo {author} {\bibfnamefont {M.}~\bibnamefont {Rzchowski}}, \bibinfo
  {author} {\bibfnamefont {J.}~\bibnamefont {Levy}},\ and\ \bibinfo {author}
  {\bibfnamefont {C.-B.}\ \bibnamefont {Eom}},\ }\bibfield  {title} {\bibinfo
  {title} {Creation of a two-dimensional electron gas at an oxide interface on
  silicon},\ }\href {https://doi.org/10.1038/ncomms1096} {\bibfield  {journal}
  {\bibinfo  {journal} {Nature Communications}\ }\textbf {\bibinfo {volume}
  {1}},\ \bibinfo {pages} {94} (\bibinfo {year} {2010})}\BibitemShut {NoStop}%
\bibitem [{\citenamefont {Popok}\ \emph {et~al.}(2020)\citenamefont {Popok},
  \citenamefont {Caban}, \citenamefont {Michalowski}, \citenamefont {Thorpe},
  \citenamefont {Feldman},\ and\ \citenamefont {Pedersen}}]{popo:20}%
  \BibitemOpen
  \bibfield  {author} {\bibinfo {author} {\bibfnamefont {V.~N.}\ \bibnamefont
  {Popok}}, \bibinfo {author} {\bibfnamefont {P.~A.}\ \bibnamefont {Caban}},
  \bibinfo {author} {\bibfnamefont {P.~P.}\ \bibnamefont {Michalowski}},
  \bibinfo {author} {\bibfnamefont {R.}~\bibnamefont {Thorpe}}, \bibinfo
  {author} {\bibfnamefont {L.~C.}\ \bibnamefont {Feldman}},\ and\ \bibinfo
  {author} {\bibfnamefont {K.}~\bibnamefont {Pedersen}},\ }\bibfield  {title}
  {\bibinfo {title} {{Two-dimensional electron gas at the AlGaN/GaN interface:
  Layer thickness dependence}},\ }\href {https://doi.org/10.1063/1.5142766}
  {\bibfield  {journal} {\bibinfo  {journal} {Journal of Applied Physics}\
  }\textbf {\bibinfo {volume} {127}},\ \bibinfo {pages} {115703} (\bibinfo
  {year} {2020})},\ \Eprint
  {https://arxiv.org/abs/https://doi.org/10.1063/1.5142766}
  {https://doi.org/10.1063/1.5142766} \BibitemShut {NoStop}%
\bibitem [{\citenamefont {Kirichenko}\ \emph {et~al.}(2017)\citenamefont
  {Kirichenko}, \citenamefont {Stephanovich},\ and\ \citenamefont
  {Dugaev}}]{kiri:17}%
  \BibitemOpen
  \bibfield  {author} {\bibinfo {author} {\bibfnamefont {E.~V.}\ \bibnamefont
  {Kirichenko}}, \bibinfo {author} {\bibfnamefont {V.~A.}\ \bibnamefont
  {Stephanovich}},\ and\ \bibinfo {author} {\bibfnamefont {V.~K.}\ \bibnamefont
  {Dugaev}},\ }\bibfield  {title} {\bibinfo {title} {{Conductivity of the
  two-dimensional electron gas at ${\mathrm{LaAlO}}_{3}/{\mathrm{SrTiO}}_{3}$
  interface}},\ }\href {https://doi.org/10.1103/PhysRevB.95.085305} {\bibfield
  {journal} {\bibinfo  {journal} {Phys. Rev. B}\ }\textbf {\bibinfo {volume}
  {95}},\ \bibinfo {pages} {085305} (\bibinfo {year} {2017})}\BibitemShut
  {NoStop}%
\bibitem [{\citenamefont {Pesquera}\ \emph {et~al.}(2014)\citenamefont
  {Pesquera}, \citenamefont {Scigaj}, \citenamefont {Gargiani}, \citenamefont
  {Barla}, \citenamefont {Herrero-Mart\'{\i}n}, \citenamefont {Pellegrin},
  \citenamefont {Valvidares}, \citenamefont {G\'azquez}, \citenamefont
  {Varela}, \citenamefont {Dix}, \citenamefont {Fontcuberta}, \citenamefont
  {S\'anchez},\ and\ \citenamefont {Herranz}}]{pesq:14}%
  \BibitemOpen
  \bibfield  {author} {\bibinfo {author} {\bibfnamefont {D.}~\bibnamefont
  {Pesquera}}, \bibinfo {author} {\bibfnamefont {M.}~\bibnamefont {Scigaj}},
  \bibinfo {author} {\bibfnamefont {P.}~\bibnamefont {Gargiani}}, \bibinfo
  {author} {\bibfnamefont {A.}~\bibnamefont {Barla}}, \bibinfo {author}
  {\bibfnamefont {J.}~\bibnamefont {Herrero-Mart\'{\i}n}}, \bibinfo {author}
  {\bibfnamefont {E.}~\bibnamefont {Pellegrin}}, \bibinfo {author}
  {\bibfnamefont {S.~M.}\ \bibnamefont {Valvidares}}, \bibinfo {author}
  {\bibfnamefont {J.}~\bibnamefont {G\'azquez}}, \bibinfo {author}
  {\bibfnamefont {M.}~\bibnamefont {Varela}}, \bibinfo {author} {\bibfnamefont
  {N.}~\bibnamefont {Dix}}, \bibinfo {author} {\bibfnamefont {J.}~\bibnamefont
  {Fontcuberta}}, \bibinfo {author} {\bibfnamefont {F.}~\bibnamefont
  {S\'anchez}},\ and\ \bibinfo {author} {\bibfnamefont {G.}~\bibnamefont
  {Herranz}},\ }\bibfield  {title} {\bibinfo {title} {{Two-Dimensional Electron
  Gases at LaAlO$_{3}$/SrTiO$_{3}$ Interfaces: Orbital Symmetry and Hierarchy
  Engineered by Crystal Orientation}},\ }\href
  {https://doi.org/10.1103/PhysRevLett.113.156802} {\bibfield  {journal}
  {\bibinfo  {journal} {Phys. Rev. Lett.}\ }\textbf {\bibinfo {volume} {113}},\
  \bibinfo {pages} {156802} (\bibinfo {year} {2014})}\BibitemShut {NoStop}%
\bibitem [{\citenamefont {Stemmer}\ and\ \citenamefont
  {James~Allen}(2014)}]{stem:14}%
  \BibitemOpen
  \bibfield  {author} {\bibinfo {author} {\bibfnamefont {S.}~\bibnamefont
  {Stemmer}}\ and\ \bibinfo {author} {\bibfnamefont {S.}~\bibnamefont
  {James~Allen}},\ }\bibfield  {title} {\bibinfo {title} {Two-dimensional
  electron gases at complex oxide interfaces},\ }\href
  {https://doi.org/10.1146/annurev-matsci-070813-113552} {\bibfield  {journal}
  {\bibinfo  {journal} {Annual Review of Materials Research}\ }\textbf
  {\bibinfo {volume} {44}},\ \bibinfo {pages} {151} (\bibinfo {year} {2014})},\
  \Eprint
  {https://arxiv.org/abs/https://doi.org/10.1146/annurev-matsci-070813-113552}
  {https://doi.org/10.1146/annurev-matsci-070813-113552} \BibitemShut {NoStop}%
\bibitem [{\citenamefont {Bello}\ \emph {et~al.}(1981)\citenamefont {Bello},
  \citenamefont {Levin}, \citenamefont {Shklovskii},\ and\ \citenamefont
  {Efros}}]{bell:81}%
  \BibitemOpen
  \bibfield  {author} {\bibinfo {author} {\bibfnamefont {M.}~\bibnamefont
  {Bello}}, \bibinfo {author} {\bibfnamefont {E.}~\bibnamefont {Levin}},
  \bibinfo {author} {\bibfnamefont {B.}~\bibnamefont {Shklovskii}},\ and\
  \bibinfo {author} {\bibfnamefont {A.}~\bibnamefont {Efros}},\ }\bibfield
  {title} {\bibinfo {title} {Density of localized states in the surface
  impurity band of a metal-insulator-semiconductor structure},\ }\href@noop {}
  {\bibfield  {journal} {\bibinfo  {journal} {JETP}\ }\textbf {\bibinfo
  {volume} {53}},\ \bibinfo {pages} {822} (\bibinfo {year} {1981})}\BibitemShut
  {NoStop}%
\bibitem [{\citenamefont {Kopp}\ and\ \citenamefont
  {Mannhart}(2009)}]{kopp:09}%
  \BibitemOpen
  \bibfield  {author} {\bibinfo {author} {\bibfnamefont {T.}~\bibnamefont
  {Kopp}}\ and\ \bibinfo {author} {\bibfnamefont {J.}~\bibnamefont
  {Mannhart}},\ }\bibfield  {title} {\bibinfo {title} {Calculation of the
  capacitances of conductors -- perspectives for the optimization of electronic
  devices},\ }\href {https://doi.org/10.1063/1.3197246} {\bibfield  {journal}
  {\bibinfo  {journal} {Journal of Applied Physics}\ }\textbf {\bibinfo
  {volume} {106}},\ \bibinfo {pages} {064504 } (\bibinfo {year}
  {2009})}\BibitemShut {NoStop}%
\bibitem [{\citenamefont {Tanatar}\ and\ \citenamefont
  {Ceperley}(1989)}]{tana:89}%
  \BibitemOpen
  \bibfield  {author} {\bibinfo {author} {\bibfnamefont {B.}~\bibnamefont
  {Tanatar}}\ and\ \bibinfo {author} {\bibfnamefont {D.}~\bibnamefont
  {Ceperley}},\ }\bibfield  {title} {\bibinfo {title} {Ground state of the
  two-dimensional electron gas},\ }\href
  {https://doi.org/10.1103/PhysRevB.39.5005} {\bibfield  {journal} {\bibinfo
  {journal} {Physical review. B, Condensed matter}\ }\textbf {\bibinfo {volume}
  {39}},\ \bibinfo {pages} {5005} (\bibinfo {year} {1989})}\BibitemShut
  {NoStop}%
\bibitem [{\citenamefont {Schakel}(2001)}]{scha:01}%
  \BibitemOpen
  \bibfield  {author} {\bibinfo {author} {\bibfnamefont {A.~M.}\ \bibnamefont
  {Schakel}},\ }\bibfield  {title} {\bibinfo {title} {Ground state of electron
  gases at negative compressibility},\ }\href@noop {} {\bibfield  {journal}
  {\bibinfo  {journal} {Physical Review B}\ }\textbf {\bibinfo {volume} {64}},\
  \bibinfo {pages} {245101} (\bibinfo {year} {2001})}\BibitemShut {NoStop}%
\bibitem [{\citenamefont {Skinner}\ and\ \citenamefont
  {Shklovskii}(2010)}]{skin:10}%
  \BibitemOpen
  \bibfield  {author} {\bibinfo {author} {\bibfnamefont {B.}~\bibnamefont
  {Skinner}}\ and\ \bibinfo {author} {\bibfnamefont {B.~I.}\ \bibnamefont
  {Shklovskii}},\ }\bibfield  {title} {\bibinfo {title} {Anomalously large
  capacitance of a plane capacitor with a two-dimensional electron gas},\
  }\href {https://doi.org/10.1103/PhysRevB.82.155111} {\bibfield  {journal}
  {\bibinfo  {journal} {Phys. Rev. B}\ }\textbf {\bibinfo {volume} {82}},\
  \bibinfo {pages} {155111} (\bibinfo {year} {2010})}\BibitemShut {NoStop}%
\bibitem [{\citenamefont {Kravchenko}\ \emph {et~al.}(1990)\citenamefont
  {Kravchenko}, \citenamefont {Rinberg}, \citenamefont {Semenchinsky},\ and\
  \citenamefont {Pudalov}}]{krav:90}%
  \BibitemOpen
  \bibfield  {author} {\bibinfo {author} {\bibfnamefont {S.~V.}\ \bibnamefont
  {Kravchenko}}, \bibinfo {author} {\bibfnamefont {D.~A.}\ \bibnamefont
  {Rinberg}}, \bibinfo {author} {\bibfnamefont {S.~G.}\ \bibnamefont
  {Semenchinsky}},\ and\ \bibinfo {author} {\bibfnamefont {V.~M.}\ \bibnamefont
  {Pudalov}},\ }\bibfield  {title} {\bibinfo {title} {Evidence for the
  influence of electron-electron interaction on the chemical potential of the
  two-dimensional electron gas},\ }\href
  {https://doi.org/10.1103/PhysRevB.42.3741} {\bibfield  {journal} {\bibinfo
  {journal} {Phys. Rev. B}\ }\textbf {\bibinfo {volume} {42}},\ \bibinfo
  {pages} {3741} (\bibinfo {year} {1990})}\BibitemShut {NoStop}%
\bibitem [{\citenamefont {Eisenstein}\ \emph {et~al.}(1992)\citenamefont
  {Eisenstein}, \citenamefont {Pfeiffer},\ and\ \citenamefont
  {West}}]{eise:92}%
  \BibitemOpen
  \bibfield  {author} {\bibinfo {author} {\bibfnamefont {J.~P.}\ \bibnamefont
  {Eisenstein}}, \bibinfo {author} {\bibfnamefont {L.~N.}\ \bibnamefont
  {Pfeiffer}},\ and\ \bibinfo {author} {\bibfnamefont {K.~W.}\ \bibnamefont
  {West}},\ }\bibfield  {title} {\bibinfo {title} {Negative compressibility of
  interacting two-dimensional electron and quasiparticle gases},\ }\href
  {https://doi.org/10.1103/PhysRevLett.68.674} {\bibfield  {journal} {\bibinfo
  {journal} {Phys. Rev. Lett.}\ }\textbf {\bibinfo {volume} {68}},\ \bibinfo
  {pages} {674} (\bibinfo {year} {1992})}\BibitemShut {NoStop}%
\bibitem [{\citenamefont {Eisenstein}\ \emph {et~al.}(1994)\citenamefont
  {Eisenstein}, \citenamefont {Pfeiffer},\ and\ \citenamefont
  {West}}]{eise:94}%
  \BibitemOpen
  \bibfield  {author} {\bibinfo {author} {\bibfnamefont {J.~P.}\ \bibnamefont
  {Eisenstein}}, \bibinfo {author} {\bibfnamefont {L.~N.}\ \bibnamefont
  {Pfeiffer}},\ and\ \bibinfo {author} {\bibfnamefont {K.~W.}\ \bibnamefont
  {West}},\ }\bibfield  {title} {\bibinfo {title} {Compressibility of the
  two-dimensional electron gas: Measurements of the zero-field exchange energy
  and fractional quantum hall gap},\ }\href
  {https://doi.org/10.1103/PhysRevB.50.1760} {\bibfield  {journal} {\bibinfo
  {journal} {Phys. Rev. B}\ }\textbf {\bibinfo {volume} {50}},\ \bibinfo
  {pages} {1760} (\bibinfo {year} {1994})}\BibitemShut {NoStop}%
\bibitem [{\citenamefont {Dultz}\ and\ \citenamefont {Jiang}(2000)}]{dult:00}%
  \BibitemOpen
  \bibfield  {author} {\bibinfo {author} {\bibfnamefont {S.~C.}\ \bibnamefont
  {Dultz}}\ and\ \bibinfo {author} {\bibfnamefont {H.~W.}\ \bibnamefont
  {Jiang}},\ }\bibfield  {title} {\bibinfo {title} {Thermodynamic signature of
  a two-dimensional metal-insulator transition},\ }\href
  {https://doi.org/10.1103/PhysRevLett.84.4689} {\bibfield  {journal} {\bibinfo
   {journal} {Phys. Rev. Lett.}\ }\textbf {\bibinfo {volume} {84}},\ \bibinfo
  {pages} {4689} (\bibinfo {year} {2000})}\BibitemShut {NoStop}%
\bibitem [{\citenamefont {Kusminskiy}\ \emph {et~al.}(2008)\citenamefont
  {Kusminskiy}, \citenamefont {Nilsson}, \citenamefont {Campbell},\ and\
  \citenamefont {Castro~Neto}}]{kusm:08}%
  \BibitemOpen
  \bibfield  {author} {\bibinfo {author} {\bibfnamefont {S.~V.}\ \bibnamefont
  {Kusminskiy}}, \bibinfo {author} {\bibfnamefont {J.}~\bibnamefont {Nilsson}},
  \bibinfo {author} {\bibfnamefont {D.~K.}\ \bibnamefont {Campbell}},\ and\
  \bibinfo {author} {\bibfnamefont {A.~H.}\ \bibnamefont {Castro~Neto}},\
  }\bibfield  {title} {\bibinfo {title} {Electronic compressibility of a
  graphene bilayer},\ }\href {https://doi.org/10.1103/PhysRevLett.100.106805}
  {\bibfield  {journal} {\bibinfo  {journal} {Phys. Rev. Lett.}\ }\textbf
  {\bibinfo {volume} {100}},\ \bibinfo {pages} {106805} (\bibinfo {year}
  {2008})}\BibitemShut {NoStop}%
\bibitem [{\citenamefont {Junquera}\ \emph {et~al.}(2019)\citenamefont
  {Junquera}, \citenamefont {Garc\'{\i}a-Fern\'andez},\ and\ \citenamefont
  {Stengel}}]{junq:19}%
  \BibitemOpen
  \bibfield  {author} {\bibinfo {author} {\bibfnamefont {J.}~\bibnamefont
  {Junquera}}, \bibinfo {author} {\bibfnamefont {P.}~\bibnamefont
  {Garc\'{\i}a-Fern\'andez}},\ and\ \bibinfo {author} {\bibfnamefont
  {M.}~\bibnamefont {Stengel}},\ }\bibfield  {title} {\bibinfo {title}
  {Mechanisms to enhance the capacitance beyond the classical limits in
  capacitors with free-electron-like electrodes},\ }\href
  {https://doi.org/10.1103/PhysRevB.99.235127} {\bibfield  {journal} {\bibinfo
  {journal} {Phys. Rev. B}\ }\textbf {\bibinfo {volume} {99}},\ \bibinfo
  {pages} {235127} (\bibinfo {year} {2019})}\BibitemShut {NoStop}%
\bibitem [{\citenamefont {Wen}\ \emph {et~al.}(2020)\citenamefont {Wen},
  \citenamefont {Zhao}, \citenamefont {Hong}, \citenamefont {Song},\ and\
  \citenamefont {He}}]{wen:20}%
  \BibitemOpen
  \bibfield  {author} {\bibinfo {author} {\bibfnamefont {W.}~\bibnamefont
  {Wen}}, \bibinfo {author} {\bibfnamefont {G.}~\bibnamefont {Zhao}}, \bibinfo
  {author} {\bibfnamefont {C.}~\bibnamefont {Hong}}, \bibinfo {author}
  {\bibfnamefont {Z.}~\bibnamefont {Song}},\ and\ \bibinfo {author}
  {\bibfnamefont {R.-H.}\ \bibnamefont {He}},\ }\bibfield  {title} {\bibinfo
  {title} {3d negative electronic compressibility as a new emergent
  phenomenon},\ }\href {https://doi.org/10.1007/s10948-019-05325-z} {\bibfield
  {journal} {\bibinfo  {journal} {Journal of Superconductivity and Novel
  Magnetism}\ }\textbf {\bibinfo {volume} {33}},\ \bibinfo {pages} {229}
  (\bibinfo {year} {2020})}\BibitemShut {NoStop}%
\bibitem [{\citenamefont {Li}\ \emph {et~al.}(2011)\citenamefont {Li},
  \citenamefont {Richter}, \citenamefont {Paetel}, \citenamefont {Kopp},
  \citenamefont {Mannhart},\ and\ \citenamefont {Ashoori}}]{li:11}%
  \BibitemOpen
  \bibfield  {author} {\bibinfo {author} {\bibfnamefont {L.}~\bibnamefont
  {Li}}, \bibinfo {author} {\bibfnamefont {C.}~\bibnamefont {Richter}},
  \bibinfo {author} {\bibfnamefont {S.}~\bibnamefont {Paetel}}, \bibinfo
  {author} {\bibfnamefont {T.}~\bibnamefont {Kopp}}, \bibinfo {author}
  {\bibfnamefont {J.}~\bibnamefont {Mannhart}},\ and\ \bibinfo {author}
  {\bibfnamefont {R.}~\bibnamefont {Ashoori}},\ }\bibfield  {title} {\bibinfo
  {title} {Very large capacitance enhancement in a two-dimensional electron
  system},\ }\href {https://doi.org/10.1126/science.1204168} {\bibfield
  {journal} {\bibinfo  {journal} {Science (New York, N.Y.)}\ }\textbf {\bibinfo
  {volume} {332}},\ \bibinfo {pages} {825} (\bibinfo {year}
  {2011})}\BibitemShut {NoStop}%
\bibitem [{\citenamefont {Tinkl}\ \emph {et~al.}(2012)\citenamefont {Tinkl},
  \citenamefont {Breitschaft}, \citenamefont {Richter},\ and\ \citenamefont
  {Mannhart}}]{tink:12}%
  \BibitemOpen
  \bibfield  {author} {\bibinfo {author} {\bibfnamefont {V.}~\bibnamefont
  {Tinkl}}, \bibinfo {author} {\bibfnamefont {M.}~\bibnamefont {Breitschaft}},
  \bibinfo {author} {\bibfnamefont {C.}~\bibnamefont {Richter}},\ and\ \bibinfo
  {author} {\bibfnamefont {J.}~\bibnamefont {Mannhart}},\ }\bibfield  {title}
  {\bibinfo {title} {{Large negative electronic compressibility of
  LaAlO$_{3}$-SrTiO$_{3}$ interfaces with ultrathin LaAlO$_{3}$ layers}},\
  }\href {https://doi.org/10.1103/PhysRevB.86.075116} {\bibfield  {journal}
  {\bibinfo  {journal} {Physical Review B}\ }\textbf {\bibinfo {volume} {86}},\
  \bibinfo {pages} {075116} (\bibinfo {year} {2012})}\BibitemShut {NoStop}%
\bibitem [{\citenamefont {Smink}\ \emph {et~al.}(2017)\citenamefont {Smink},
  \citenamefont {de~Boer}, \citenamefont {Stehno}, \citenamefont {Brinkman},
  \citenamefont {van~der Wiel},\ and\ \citenamefont {Hilgenkamp}}]{smin:17}%
  \BibitemOpen
  \bibfield  {author} {\bibinfo {author} {\bibfnamefont {A.~E.~M.}\
  \bibnamefont {Smink}}, \bibinfo {author} {\bibfnamefont {J.~C.}\ \bibnamefont
  {de~Boer}}, \bibinfo {author} {\bibfnamefont {M.~P.}\ \bibnamefont {Stehno}},
  \bibinfo {author} {\bibfnamefont {A.}~\bibnamefont {Brinkman}}, \bibinfo
  {author} {\bibfnamefont {W.~G.}\ \bibnamefont {van~der Wiel}},\ and\ \bibinfo
  {author} {\bibfnamefont {H.}~\bibnamefont {Hilgenkamp}},\ }\bibfield  {title}
  {\bibinfo {title} {{Gate-Tunable Band Structure of the
  LaAlO$_{3}$-SrTiO$_{3}$ Interface}},\ }\href
  {https://doi.org/10.1103/PhysRevLett.118.106401} {\bibfield  {journal}
  {\bibinfo  {journal} {Phys. Rev. Lett.}\ }\textbf {\bibinfo {volume} {118}},\
  \bibinfo {pages} {106401} (\bibinfo {year} {2017})}\BibitemShut {NoStop}%
\bibitem [{\citenamefont {Ohtomo}\ and\ \citenamefont {Hwang}(2006)}]{ohto:04}%
  \BibitemOpen
  \bibfield  {author} {\bibinfo {author} {\bibfnamefont {A.}~\bibnamefont
  {Ohtomo}}\ and\ \bibinfo {author} {\bibfnamefont {H.}~\bibnamefont {Hwang}},\
  }\bibfield  {title} {\bibinfo {title} {{Corrigendum: A high-mobility electron
  gas at the LaAlO3/SrTiO3 heterointerface}},\ }\href
  {https://doi.org/10.1038/nature04773} {\bibfield  {journal} {\bibinfo
  {journal} {Nature}\ }\textbf {\bibinfo {volume} {441}},\ \bibinfo {pages}
  {120} (\bibinfo {year} {2006})}\BibitemShut {NoStop}%
\bibitem [{\citenamefont {Reyren}\ \emph {et~al.}(2007)\citenamefont {Reyren},
  \citenamefont {Paetel}, \citenamefont {Caviglia}, \citenamefont {Kourkoutis},
  \citenamefont {Hammerl}, \citenamefont {Richter}, \citenamefont {Schneider},
  \citenamefont {Kopp}, \citenamefont {Rüetschi}, \citenamefont {Jaccard},
  \citenamefont {Gabay}, \citenamefont {Muller}, \citenamefont {Triscone},\
  and\ \citenamefont {Mannhart}}]{reyr:07}%
  \BibitemOpen
  \bibfield  {author} {\bibinfo {author} {\bibfnamefont {N.}~\bibnamefont
  {Reyren}}, \bibinfo {author} {\bibfnamefont {S.}~\bibnamefont {Paetel}},
  \bibinfo {author} {\bibfnamefont {A.}~\bibnamefont {Caviglia}}, \bibinfo
  {author} {\bibfnamefont {L.}~\bibnamefont {Kourkoutis}}, \bibinfo {author}
  {\bibfnamefont {G.}~\bibnamefont {Hammerl}}, \bibinfo {author} {\bibfnamefont
  {C.}~\bibnamefont {Richter}}, \bibinfo {author} {\bibfnamefont
  {C.}~\bibnamefont {Schneider}}, \bibinfo {author} {\bibfnamefont
  {T.}~\bibnamefont {Kopp}}, \bibinfo {author} {\bibfnamefont {A.-S.}\
  \bibnamefont {Rüetschi}}, \bibinfo {author} {\bibfnamefont {D.}~\bibnamefont
  {Jaccard}}, \bibinfo {author} {\bibfnamefont {M.}~\bibnamefont {Gabay}},
  \bibinfo {author} {\bibfnamefont {D.}~\bibnamefont {Muller}}, \bibinfo
  {author} {\bibfnamefont {J.-M.}\ \bibnamefont {Triscone}},\ and\ \bibinfo
  {author} {\bibfnamefont {J.}~\bibnamefont {Mannhart}},\ }\bibfield  {title}
  {\bibinfo {title} {Superconducting interfaces between insulating oxides},\
  }\href {https://doi.org/10.1126/science.1146006} {\bibfield  {journal}
  {\bibinfo  {journal} {Science (New York, N.Y.)}\ }\textbf {\bibinfo {volume}
  {317}},\ \bibinfo {pages} {1196} (\bibinfo {year} {2007})}\BibitemShut
  {NoStop}%
\bibitem [{\citenamefont {Biscaras}\ \emph {et~al.}(2012)\citenamefont
  {Biscaras}, \citenamefont {Bergeal}, \citenamefont {Hurand}, \citenamefont
  {Grosset\^ete}, \citenamefont {Rastogi}, \citenamefont {Budhani},
  \citenamefont {LeBoeuf}, \citenamefont {Proust},\ and\ \citenamefont
  {Lesueur}}]{bisc:12}%
  \BibitemOpen
  \bibfield  {author} {\bibinfo {author} {\bibfnamefont {J.}~\bibnamefont
  {Biscaras}}, \bibinfo {author} {\bibfnamefont {N.}~\bibnamefont {Bergeal}},
  \bibinfo {author} {\bibfnamefont {S.}~\bibnamefont {Hurand}}, \bibinfo
  {author} {\bibfnamefont {C.}~\bibnamefont {Grosset\^ete}}, \bibinfo {author}
  {\bibfnamefont {A.}~\bibnamefont {Rastogi}}, \bibinfo {author} {\bibfnamefont
  {R.~C.}\ \bibnamefont {Budhani}}, \bibinfo {author} {\bibfnamefont
  {D.}~\bibnamefont {LeBoeuf}}, \bibinfo {author} {\bibfnamefont
  {C.}~\bibnamefont {Proust}},\ and\ \bibinfo {author} {\bibfnamefont
  {J.}~\bibnamefont {Lesueur}},\ }\bibfield  {title} {\bibinfo {title}
  {{Two-Dimensional Superconducting Phase in LaTiO$_3$/SrTiO$_3$
  Heterostructures Induced by High-Mobility Carrier Doping}},\ }\href
  {https://doi.org/10.1103/PhysRevLett.108.247004} {\bibfield  {journal}
  {\bibinfo  {journal} {Phys. Rev. Lett.}\ }\textbf {\bibinfo {volume} {108}},\
  \bibinfo {pages} {247004} (\bibinfo {year} {2012})}\BibitemShut {NoStop}%
\bibitem [{\citenamefont {Caviglia}\ \emph {et~al.}(2009)\citenamefont
  {Caviglia}, \citenamefont {Gariglio}, \citenamefont {Reyren}, \citenamefont
  {Jaccard}, \citenamefont {Schneider}, \citenamefont {Gabay}, \citenamefont
  {Paetel}, \citenamefont {Hammerl}, \citenamefont {Mannhart},\ and\
  \citenamefont {Triscone}}]{cavi:08}%
  \BibitemOpen
  \bibfield  {author} {\bibinfo {author} {\bibfnamefont {A.}~\bibnamefont
  {Caviglia}}, \bibinfo {author} {\bibfnamefont {S.}~\bibnamefont {Gariglio}},
  \bibinfo {author} {\bibfnamefont {N.}~\bibnamefont {Reyren}}, \bibinfo
  {author} {\bibfnamefont {D.}~\bibnamefont {Jaccard}}, \bibinfo {author}
  {\bibfnamefont {T.}~\bibnamefont {Schneider}}, \bibinfo {author}
  {\bibfnamefont {M.}~\bibnamefont {Gabay}}, \bibinfo {author} {\bibfnamefont
  {S.}~\bibnamefont {Paetel}}, \bibinfo {author} {\bibfnamefont
  {G.}~\bibnamefont {Hammerl}}, \bibinfo {author} {\bibfnamefont
  {J.}~\bibnamefont {Mannhart}},\ and\ \bibinfo {author} {\bibfnamefont
  {J.-M.}\ \bibnamefont {Triscone}},\ }\bibfield  {title} {\bibinfo {title}
  {{Electric field control of the LaAlO$_3$/SrTiO$_3$ interface ground
  state}},\ }\href {https://doi.org/10.1038/nature07576} {\bibfield  {journal}
  {\bibinfo  {journal} {Nature}\ }\textbf {\bibinfo {volume} {456}},\ \bibinfo
  {pages} {624} (\bibinfo {year} {2009})}\BibitemShut {NoStop}%
\bibitem [{\citenamefont {Bell}\ \emph {et~al.}(2009)\citenamefont {Bell},
  \citenamefont {Harashima}, \citenamefont {Kozuka}, \citenamefont {Kim},
  \citenamefont {Kim}, \citenamefont {Hikita},\ and\ \citenamefont
  {Hwang}}]{bell:09}%
  \BibitemOpen
  \bibfield  {author} {\bibinfo {author} {\bibfnamefont {C.}~\bibnamefont
  {Bell}}, \bibinfo {author} {\bibfnamefont {S.}~\bibnamefont {Harashima}},
  \bibinfo {author} {\bibfnamefont {Y.}~\bibnamefont {Kozuka}}, \bibinfo
  {author} {\bibfnamefont {M.}~\bibnamefont {Kim}}, \bibinfo {author}
  {\bibfnamefont {B.}~\bibnamefont {Kim}}, \bibinfo {author} {\bibfnamefont
  {Y.}~\bibnamefont {Hikita}},\ and\ \bibinfo {author} {\bibfnamefont
  {H.}~\bibnamefont {Hwang}},\ }\bibfield  {title} {\bibinfo {title} {{Dominant
  Mobility Modulation by the Electric Field Effect at the LaAlO$_3$ / SrTiO$_3$
  Interface}},\ }\href {https://doi.org/10.1103/PhysRevLett.103.226802}
  {\bibfield  {journal} {\bibinfo  {journal} {Physical Review Letters}\
  }\textbf {\bibinfo {volume} {103}},\ \bibinfo {pages} {226802} (\bibinfo
  {year} {2009})}\BibitemShut {NoStop}%
\bibitem [{\citenamefont {Fernandes}\ \emph {et~al.}(2013)\citenamefont
  {Fernandes}, \citenamefont {Haraldsen}, \citenamefont {W\"olfle},\ and\
  \citenamefont {Balatsky}}]{fern:13}%
  \BibitemOpen
  \bibfield  {author} {\bibinfo {author} {\bibfnamefont {R.~M.}\ \bibnamefont
  {Fernandes}}, \bibinfo {author} {\bibfnamefont {J.~T.}\ \bibnamefont
  {Haraldsen}}, \bibinfo {author} {\bibfnamefont {P.}~\bibnamefont
  {W\"olfle}},\ and\ \bibinfo {author} {\bibfnamefont {A.~V.}\ \bibnamefont
  {Balatsky}},\ }\bibfield  {title} {\bibinfo {title} {{Two-band
  superconductivity in doped SrTiO$_{3}$ films and interfaces}},\ }\href
  {https://doi.org/10.1103/PhysRevB.87.014510} {\bibfield  {journal} {\bibinfo
  {journal} {Phys. Rev. B}\ }\textbf {\bibinfo {volume} {87}},\ \bibinfo
  {pages} {014510} (\bibinfo {year} {2013})}\BibitemShut {NoStop}%
\bibitem [{\citenamefont {Haraldsen}\ and\ \citenamefont
  {Balatsky}(2013)}]{hara:13}%
  \BibitemOpen
  \bibfield  {author} {\bibinfo {author} {\bibfnamefont {J.}~\bibnamefont
  {Haraldsen}}\ and\ \bibinfo {author} {\bibfnamefont {A.}~\bibnamefont
  {Balatsky}},\ }\bibfield  {title} {\bibinfo {title} {Effects of
  magnetoelectric ordering due to interfacial symmetry breaking},\ }\href
  {https://doi.org/10.1080/21663831.2013.764942} {\bibfield  {journal}
  {\bibinfo  {journal} {Materials Research Letters}\ }\textbf {\bibinfo
  {volume} {1}},\ \bibinfo {pages} {39} (\bibinfo {year} {2013})}\BibitemShut
  {NoStop}%
\bibitem [{\citenamefont {Singh}\ \emph {et~al.}(2018)\citenamefont {Singh},
  \citenamefont {Jouan}, \citenamefont {Benfatto}, \citenamefont {Cou{\"e}do},
  \citenamefont {Kumar}, \citenamefont {Dogra}, \citenamefont {Budhani},
  \citenamefont {Caprara}, \citenamefont {Grilli}, \citenamefont {Lesne} \emph
  {et~al.}}]{sing:18}%
  \BibitemOpen
  \bibfield  {author} {\bibinfo {author} {\bibfnamefont {G.}~\bibnamefont
  {Singh}}, \bibinfo {author} {\bibfnamefont {A.}~\bibnamefont {Jouan}},
  \bibinfo {author} {\bibfnamefont {L.}~\bibnamefont {Benfatto}}, \bibinfo
  {author} {\bibfnamefont {F.}~\bibnamefont {Cou{\"e}do}}, \bibinfo {author}
  {\bibfnamefont {P.}~\bibnamefont {Kumar}}, \bibinfo {author} {\bibfnamefont
  {A.}~\bibnamefont {Dogra}}, \bibinfo {author} {\bibfnamefont
  {R.}~\bibnamefont {Budhani}}, \bibinfo {author} {\bibfnamefont
  {S.}~\bibnamefont {Caprara}}, \bibinfo {author} {\bibfnamefont
  {M.}~\bibnamefont {Grilli}}, \bibinfo {author} {\bibfnamefont
  {E.}~\bibnamefont {Lesne}}, \emph {et~al.},\ }\bibfield  {title} {\bibinfo
  {title} {Competition between electron pairing and phase coherence in
  superconducting interfaces},\ }\href@noop {} {\bibfield  {journal} {\bibinfo
  {journal} {Nature communications}\ }\textbf {\bibinfo {volume} {9}},\
  \bibinfo {pages} {1} (\bibinfo {year} {2018})}\BibitemShut {NoStop}%
\bibitem [{\citenamefont {Boschker}\ \emph {et~al.}(2015)\citenamefont
  {Boschker}, \citenamefont {Richter}, \citenamefont {Fillis-Tsirakis},
  \citenamefont {Schneider},\ and\ \citenamefont {Mannhart}}]{bosc:15}%
  \BibitemOpen
  \bibfield  {author} {\bibinfo {author} {\bibfnamefont {H.}~\bibnamefont
  {Boschker}}, \bibinfo {author} {\bibfnamefont {C.}~\bibnamefont {Richter}},
  \bibinfo {author} {\bibfnamefont {E.}~\bibnamefont {Fillis-Tsirakis}},
  \bibinfo {author} {\bibfnamefont {C.~W.}\ \bibnamefont {Schneider}},\ and\
  \bibinfo {author} {\bibfnamefont {J.}~\bibnamefont {Mannhart}},\ }\bibfield
  {title} {\bibinfo {title} {{Electron—phonon Coupling and the
  Superconducting Phase Diagram of the LaAlO$_3$—SrTiO$_3$ Interface}},\
  }\href@noop {} {\bibfield  {journal} {\bibinfo  {journal} {Scientific
  reports}\ }\textbf {\bibinfo {volume} {5}},\ \bibinfo {pages} {12309}
  (\bibinfo {year} {2015})}\BibitemShut {NoStop}%
\bibitem [{\citenamefont {Dunnett}\ \emph {et~al.}(2018)\citenamefont
  {Dunnett}, \citenamefont {Narayan}, \citenamefont {Spaldin},\ and\
  \citenamefont {Balatsky}}]{dunn:18}%
  \BibitemOpen
  \bibfield  {author} {\bibinfo {author} {\bibfnamefont {K.}~\bibnamefont
  {Dunnett}}, \bibinfo {author} {\bibfnamefont {A.}~\bibnamefont {Narayan}},
  \bibinfo {author} {\bibfnamefont {N.~A.}\ \bibnamefont {Spaldin}},\ and\
  \bibinfo {author} {\bibfnamefont {A.~V.}\ \bibnamefont {Balatsky}},\
  }\bibfield  {title} {\bibinfo {title} {Strain and ferroelectric soft-mode
  induced superconductivity in strontium titanate},\ }\href
  {https://doi.org/10.1103/PhysRevB.97.144506} {\bibfield  {journal} {\bibinfo
  {journal} {Phys. Rev. B}\ }\textbf {\bibinfo {volume} {97}},\ \bibinfo
  {pages} {144506} (\bibinfo {year} {2018})}\BibitemShut {NoStop}%
\bibitem [{\citenamefont {Herrera}\ \emph {et~al.}(2019)\citenamefont
  {Herrera}, \citenamefont {Cerbin}, \citenamefont {Jayakody}, \citenamefont
  {Dunnett}, \citenamefont {Balatsky},\ and\ \citenamefont
  {Sochnikov}}]{herr:19}%
  \BibitemOpen
  \bibfield  {author} {\bibinfo {author} {\bibfnamefont {C.}~\bibnamefont
  {Herrera}}, \bibinfo {author} {\bibfnamefont {J.}~\bibnamefont {Cerbin}},
  \bibinfo {author} {\bibfnamefont {A.}~\bibnamefont {Jayakody}}, \bibinfo
  {author} {\bibfnamefont {K.}~\bibnamefont {Dunnett}}, \bibinfo {author}
  {\bibfnamefont {A.~V.}\ \bibnamefont {Balatsky}},\ and\ \bibinfo {author}
  {\bibfnamefont {I.}~\bibnamefont {Sochnikov}},\ }\bibfield  {title} {\bibinfo
  {title} {Strain-engineered interaction of quantum polar and superconducting
  phases},\ }\href {https://doi.org/10.1103/PhysRevMaterials.3.124801}
  {\bibfield  {journal} {\bibinfo  {journal} {Phys. Rev. Materials}\ }\textbf
  {\bibinfo {volume} {3}},\ \bibinfo {pages} {124801} (\bibinfo {year}
  {2019})}\BibitemShut {NoStop}%
\bibitem [{\citenamefont {Koonce}\ \emph {et~al.}(1967)\citenamefont {Koonce},
  \citenamefont {Cohen}, \citenamefont {Schooley}, \citenamefont {Hosler},\
  and\ \citenamefont {Pfeiffer}}]{koon:67}%
  \BibitemOpen
  \bibfield  {author} {\bibinfo {author} {\bibfnamefont {C.~S.}\ \bibnamefont
  {Koonce}}, \bibinfo {author} {\bibfnamefont {M.~L.}\ \bibnamefont {Cohen}},
  \bibinfo {author} {\bibfnamefont {J.~F.}\ \bibnamefont {Schooley}}, \bibinfo
  {author} {\bibfnamefont {W.~R.}\ \bibnamefont {Hosler}},\ and\ \bibinfo
  {author} {\bibfnamefont {E.~R.}\ \bibnamefont {Pfeiffer}},\ }\bibfield
  {title} {\bibinfo {title} {Superconducting transition temperatures of
  semiconducting srti${\mathrm{o}}_{3}$},\ }\href
  {https://doi.org/10.1103/PhysRev.163.380} {\bibfield  {journal} {\bibinfo
  {journal} {Phys. Rev.}\ }\textbf {\bibinfo {volume} {163}},\ \bibinfo {pages}
  {380} (\bibinfo {year} {1967})}\BibitemShut {NoStop}%
\bibitem [{\citenamefont {Binnig}\ \emph {et~al.}(1980)\citenamefont {Binnig},
  \citenamefont {Baratoff}, \citenamefont {Hoenig},\ and\ \citenamefont
  {Bednorz}}]{binn:80}%
  \BibitemOpen
  \bibfield  {author} {\bibinfo {author} {\bibfnamefont {G.}~\bibnamefont
  {Binnig}}, \bibinfo {author} {\bibfnamefont {A.}~\bibnamefont {Baratoff}},
  \bibinfo {author} {\bibfnamefont {H.~E.}\ \bibnamefont {Hoenig}},\ and\
  \bibinfo {author} {\bibfnamefont {J.~G.}\ \bibnamefont {Bednorz}},\
  }\bibfield  {title} {\bibinfo {title} {Two-band superconductivity in nb-doped
  srti${\mathrm{o}}_{3}$},\ }\href
  {https://doi.org/10.1103/PhysRevLett.45.1352} {\bibfield  {journal} {\bibinfo
   {journal} {Phys. Rev. Lett.}\ }\textbf {\bibinfo {volume} {45}},\ \bibinfo
  {pages} {1352} (\bibinfo {year} {1980})}\BibitemShut {NoStop}%
\bibitem [{\citenamefont {Guduru}\ \emph
  {et~al.}(2013{\natexlab{a}})\citenamefont {Guduru}, \citenamefont {McCollam},
  \citenamefont {Maan}, \citenamefont {Zeitler}, \citenamefont {Wenderich},
  \citenamefont {Kruize}, \citenamefont {Brinkman}, \citenamefont {Huijben},
  \citenamefont {Koster}, \citenamefont {Blank}, \citenamefont {Rijnders},\
  and\ \citenamefont {Hilgenkamp}}]{veer:13}%
  \BibitemOpen
  \bibfield  {author} {\bibinfo {author} {\bibfnamefont {V.}~\bibnamefont
  {Guduru}}, \bibinfo {author} {\bibfnamefont {A.}~\bibnamefont {McCollam}},
  \bibinfo {author} {\bibfnamefont {J.}~\bibnamefont {Maan}}, \bibinfo {author}
  {\bibfnamefont {U.}~\bibnamefont {Zeitler}}, \bibinfo {author} {\bibfnamefont
  {S.}~\bibnamefont {Wenderich}}, \bibinfo {author} {\bibfnamefont
  {M.}~\bibnamefont {Kruize}}, \bibinfo {author} {\bibfnamefont
  {A.}~\bibnamefont {Brinkman}}, \bibinfo {author} {\bibfnamefont
  {M.}~\bibnamefont {Huijben}}, \bibinfo {author} {\bibfnamefont
  {G.}~\bibnamefont {Koster}}, \bibinfo {author} {\bibfnamefont
  {D.}~\bibnamefont {Blank}}, \bibinfo {author} {\bibfnamefont
  {G.}~\bibnamefont {Rijnders}},\ and\ \bibinfo {author} {\bibfnamefont
  {H.}~\bibnamefont {Hilgenkamp}},\ }\bibfield  {title} {\bibinfo {title}
  {{Multi-band conduction behaviour at the interface of LaAlO$_3$/SrTiO$_3$
  heterostructures}},\ }\href {https://doi.org/10.3938/jkps.63.437} {\bibfield
  {journal} {\bibinfo  {journal} {Journal of Korean Physical Society}\ }\textbf
  {\bibinfo {volume} {63}},\ \bibinfo {pages} {437} (\bibinfo {year}
  {2013}{\natexlab{a}})}\BibitemShut {NoStop}%
\bibitem [{\citenamefont {Guduru}\ \emph
  {et~al.}(2013{\natexlab{b}})\citenamefont {Guduru}, \citenamefont {Aguila},
  \citenamefont {Wenderich}, \citenamefont {Kruize}, \citenamefont {McCollam},
  \citenamefont {Christianen}, \citenamefont {Zeitler}, \citenamefont
  {Brinkman}, \citenamefont {Rijnders}, \citenamefont {Hilgenkamp},\ and\
  \citenamefont {Maan}}]{gudu:13}%
  \BibitemOpen
  \bibfield  {author} {\bibinfo {author} {\bibfnamefont {V.}~\bibnamefont
  {Guduru}}, \bibinfo {author} {\bibfnamefont {A.}~\bibnamefont {Aguila}},
  \bibinfo {author} {\bibfnamefont {S.}~\bibnamefont {Wenderich}}, \bibinfo
  {author} {\bibfnamefont {M.}~\bibnamefont {Kruize}}, \bibinfo {author}
  {\bibfnamefont {A.}~\bibnamefont {McCollam}}, \bibinfo {author}
  {\bibfnamefont {P.}~\bibnamefont {Christianen}}, \bibinfo {author}
  {\bibfnamefont {U.}~\bibnamefont {Zeitler}}, \bibinfo {author} {\bibfnamefont
  {A.}~\bibnamefont {Brinkman}}, \bibinfo {author} {\bibfnamefont
  {G.}~\bibnamefont {Rijnders}}, \bibinfo {author} {\bibfnamefont
  {H.}~\bibnamefont {Hilgenkamp}},\ and\ \bibinfo {author} {\bibfnamefont
  {J.}~\bibnamefont {Maan}},\ }\bibfield  {title} {\bibinfo {title} {{Optically
  excited multi-band conduction in LaAlO$_3$/SrTiO$_3$ heterostructures}},\
  }\href {https://doi.org/10.1063/1.4790844} {\bibfield  {journal} {\bibinfo
  {journal} {Applied Physics Letters}\ }\textbf {\bibinfo {volume} {102}},\
  \bibinfo {pages} {051604} (\bibinfo {year} {2013}{\natexlab{b}})}\BibitemShut
  {NoStop}%
\bibitem [{\citenamefont {Song}\ \emph {et~al.}(2018)\citenamefont {Song},
  \citenamefont {Ryu}, \citenamefont {Lee}, \citenamefont {Paudel},
  \citenamefont {Koch}, \citenamefont {Park}, \citenamefont {Lee},
  \citenamefont {Choi}, \citenamefont {Kim}, \citenamefont {Kim} \emph
  {et~al.}}]{song:18}%
  \BibitemOpen
  \bibfield  {author} {\bibinfo {author} {\bibfnamefont {K.}~\bibnamefont
  {Song}}, \bibinfo {author} {\bibfnamefont {S.}~\bibnamefont {Ryu}}, \bibinfo
  {author} {\bibfnamefont {H.}~\bibnamefont {Lee}}, \bibinfo {author}
  {\bibfnamefont {T.~R.}\ \bibnamefont {Paudel}}, \bibinfo {author}
  {\bibfnamefont {C.~T.}\ \bibnamefont {Koch}}, \bibinfo {author}
  {\bibfnamefont {B.}~\bibnamefont {Park}}, \bibinfo {author} {\bibfnamefont
  {J.~K.}\ \bibnamefont {Lee}}, \bibinfo {author} {\bibfnamefont {S.-Y.}\
  \bibnamefont {Choi}}, \bibinfo {author} {\bibfnamefont {Y.-M.}\ \bibnamefont
  {Kim}}, \bibinfo {author} {\bibfnamefont {J.~C.}\ \bibnamefont {Kim}}, \emph
  {et~al.},\ }\bibfield  {title} {\bibinfo {title} {Direct imaging of the
  electron liquid at oxide interfaces},\ }\href@noop {} {\bibfield  {journal}
  {\bibinfo  {journal} {Nature nanotechnology}\ }\textbf {\bibinfo {volume}
  {13}},\ \bibinfo {pages} {198} (\bibinfo {year} {2018})}\BibitemShut
  {NoStop}%
\bibitem [{\citenamefont {Joshua}\ \emph {et~al.}(2012)\citenamefont {Joshua},
  \citenamefont {Pecker}, \citenamefont {Ruhman}, \citenamefont {Altman},\ and\
  \citenamefont {Ilani}}]{josh:12}%
  \BibitemOpen
  \bibfield  {author} {\bibinfo {author} {\bibfnamefont {A.}~\bibnamefont
  {Joshua}}, \bibinfo {author} {\bibfnamefont {S.}~\bibnamefont {Pecker}},
  \bibinfo {author} {\bibfnamefont {J.}~\bibnamefont {Ruhman}}, \bibinfo
  {author} {\bibfnamefont {E.}~\bibnamefont {Altman}},\ and\ \bibinfo {author}
  {\bibfnamefont {S.}~\bibnamefont {Ilani}},\ }\bibfield  {title} {\bibinfo
  {title} {{A universal critical density underlying the physics of electrons at
  the LaAlO$_3$/SrTiO$_3$ interface}},\ }\href
  {https://doi.org/10.1038/ncomms2116} {\bibfield  {journal} {\bibinfo
  {journal} {Nature communications}\ }\textbf {\bibinfo {volume} {3}},\
  \bibinfo {pages} {1129} (\bibinfo {year} {2012})}\BibitemShut {NoStop}%
\bibitem [{\citenamefont {van Mechelen}\ \emph {et~al.}(2008)\citenamefont {van
  Mechelen}, \citenamefont {van~der Marel}, \citenamefont {Grimaldi},
  \citenamefont {Kuzmenko}, \citenamefont {Armitage}, \citenamefont {Reyren},
  \citenamefont {Hagemann},\ and\ \citenamefont {Mazin}}]{vanm:08}%
  \BibitemOpen
  \bibfield  {author} {\bibinfo {author} {\bibfnamefont {J.~L.~M.}\
  \bibnamefont {van Mechelen}}, \bibinfo {author} {\bibfnamefont
  {D.}~\bibnamefont {van~der Marel}}, \bibinfo {author} {\bibfnamefont
  {C.}~\bibnamefont {Grimaldi}}, \bibinfo {author} {\bibfnamefont {A.~B.}\
  \bibnamefont {Kuzmenko}}, \bibinfo {author} {\bibfnamefont {N.~P.}\
  \bibnamefont {Armitage}}, \bibinfo {author} {\bibfnamefont {N.}~\bibnamefont
  {Reyren}}, \bibinfo {author} {\bibfnamefont {H.}~\bibnamefont {Hagemann}},\
  and\ \bibinfo {author} {\bibfnamefont {I.~I.}\ \bibnamefont {Mazin}},\
  }\bibfield  {title} {\bibinfo {title} {{Electron-Phonon Interaction and
  Charge Carrier Mass Enhancement in ${\mathrm{SrTiO}}_{3}$}},\ }\href
  {https://doi.org/10.1103/PhysRevLett.100.226403} {\bibfield  {journal}
  {\bibinfo  {journal} {Phys. Rev. Lett.}\ }\textbf {\bibinfo {volume} {100}},\
  \bibinfo {pages} {226403} (\bibinfo {year} {2008})}\BibitemShut {NoStop}%
\bibitem [{\citenamefont {Smink}\ \emph {et~al.}(2018)\citenamefont {Smink},
  \citenamefont {Stehno}, \citenamefont {De~Boer}, \citenamefont {Brinkman},
  \citenamefont {Van Der~Wiel},\ and\ \citenamefont {Hilgenkamp}}]{smin:18}%
  \BibitemOpen
  \bibfield  {author} {\bibinfo {author} {\bibfnamefont {A.}~\bibnamefont
  {Smink}}, \bibinfo {author} {\bibfnamefont {M.}~\bibnamefont {Stehno}},
  \bibinfo {author} {\bibfnamefont {J.}~\bibnamefont {De~Boer}}, \bibinfo
  {author} {\bibfnamefont {A.}~\bibnamefont {Brinkman}}, \bibinfo {author}
  {\bibfnamefont {W.}~\bibnamefont {Van Der~Wiel}},\ and\ \bibinfo {author}
  {\bibfnamefont {H.}~\bibnamefont {Hilgenkamp}},\ }\bibfield  {title}
  {\bibinfo {title} {Correlation between superconductivity, band filling, and
  electron confinement at the laalo 3/srtio 3 interface},\ }\href@noop {}
  {\bibfield  {journal} {\bibinfo  {journal} {Physical Review B}\ }\textbf
  {\bibinfo {volume} {97}},\ \bibinfo {pages} {245113} (\bibinfo {year}
  {2018})}\BibitemShut {NoStop}%
\bibitem [{\citenamefont {Suhl}\ \emph {et~al.}(1959)\citenamefont {Suhl},
  \citenamefont {Matthias},\ and\ \citenamefont {Walker}}]{suhl:59}%
  \BibitemOpen
  \bibfield  {author} {\bibinfo {author} {\bibfnamefont {H.}~\bibnamefont
  {Suhl}}, \bibinfo {author} {\bibfnamefont {B.~T.}\ \bibnamefont {Matthias}},\
  and\ \bibinfo {author} {\bibfnamefont {L.~R.}\ \bibnamefont {Walker}},\
  }\bibfield  {title} {\bibinfo {title} {Bardeen-cooper-schrieffer theory of
  superconductivity in the case of overlapping bands},\ }\href
  {https://doi.org/10.1103/PhysRevLett.3.552} {\bibfield  {journal} {\bibinfo
  {journal} {Phys. Rev. Lett.}\ }\textbf {\bibinfo {volume} {3}},\ \bibinfo
  {pages} {552} (\bibinfo {year} {1959})}\BibitemShut {NoStop}%
\bibitem [{\citenamefont {Radebaugh}\ and\ \citenamefont
  {Keesom}(1966)}]{rade:66}%
  \BibitemOpen
  \bibfield  {author} {\bibinfo {author} {\bibfnamefont {R.}~\bibnamefont
  {Radebaugh}}\ and\ \bibinfo {author} {\bibfnamefont {P.~H.}\ \bibnamefont
  {Keesom}},\ }\bibfield  {title} {\bibinfo {title} {Low-temperature
  thermodynamic properties of vanadium. i. superconducting and normal states},\
  }\href {https://doi.org/10.1103/PhysRev.149.209} {\bibfield  {journal}
  {\bibinfo  {journal} {Phys. Rev.}\ }\textbf {\bibinfo {volume} {149}},\
  \bibinfo {pages} {209} (\bibinfo {year} {1966})}\BibitemShut {NoStop}%
\bibitem [{\citenamefont {Shen}\ \emph {et~al.}(1965)\citenamefont {Shen},
  \citenamefont {Senozan},\ and\ \citenamefont {Phillips}}]{shen:65}%
  \BibitemOpen
  \bibfield  {author} {\bibinfo {author} {\bibfnamefont {L.~Y.~L.}\
  \bibnamefont {Shen}}, \bibinfo {author} {\bibfnamefont {N.~M.}\ \bibnamefont
  {Senozan}},\ and\ \bibinfo {author} {\bibfnamefont {N.~E.}\ \bibnamefont
  {Phillips}},\ }\bibfield  {title} {\bibinfo {title} {Evidence for two energy
  gaps in high-purity superconducting nb, ta, and v},\ }\href
  {https://doi.org/10.1103/PhysRevLett.14.1025} {\bibfield  {journal} {\bibinfo
   {journal} {Phys. Rev. Lett.}\ }\textbf {\bibinfo {volume} {14}},\ \bibinfo
  {pages} {1025} (\bibinfo {year} {1965})}\BibitemShut {NoStop}%
\bibitem [{\citenamefont {Bennett}(1965)}]{benn:65}%
  \BibitemOpen
  \bibfield  {author} {\bibinfo {author} {\bibfnamefont {A.~J.}\ \bibnamefont
  {Bennett}},\ }\bibfield  {title} {\bibinfo {title} {Theory of the anisotropic
  energy gap in superconducting lead},\ }\href
  {https://doi.org/10.1103/PhysRev.140.A1902} {\bibfield  {journal} {\bibinfo
  {journal} {Phys. Rev.}\ }\textbf {\bibinfo {volume} {140}},\ \bibinfo {pages}
  {A1902} (\bibinfo {year} {1965})}\BibitemShut {NoStop}%
\bibitem [{\citenamefont {Short}\ and\ \citenamefont {Wolfe}(2000)}]{shor:00}%
  \BibitemOpen
  \bibfield  {author} {\bibinfo {author} {\bibfnamefont {J.~D.}\ \bibnamefont
  {Short}}\ and\ \bibinfo {author} {\bibfnamefont {J.~P.}\ \bibnamefont
  {Wolfe}},\ }\bibfield  {title} {\bibinfo {title} {{Evidence for Large Gap
  Anisotropy in Superconducting Pb from Phonon Imaging}},\ }\href
  {https://doi.org/10.1103/PhysRevLett.85.5198} {\bibfield  {journal} {\bibinfo
   {journal} {Phys. Rev. Lett.}\ }\textbf {\bibinfo {volume} {85}},\ \bibinfo
  {pages} {5198} (\bibinfo {year} {2000})}\BibitemShut {NoStop}%
\bibitem [{\citenamefont {Tomlinson}\ and\ \citenamefont
  {Carbotte}(1976)}]{toml:75}%
  \BibitemOpen
  \bibfield  {author} {\bibinfo {author} {\bibfnamefont {P.~G.}\ \bibnamefont
  {Tomlinson}}\ and\ \bibinfo {author} {\bibfnamefont {J.~P.}\ \bibnamefont
  {Carbotte}},\ }\bibfield  {title} {\bibinfo {title} {{Anisotropic
  superconducting energy gap in Pb}},\ }\href
  {https://doi.org/10.1103/PhysRevB.13.4738} {\bibfield  {journal} {\bibinfo
  {journal} {Phys. Rev. B}\ }\textbf {\bibinfo {volume} {13}},\ \bibinfo
  {pages} {4738} (\bibinfo {year} {1976})}\BibitemShut {NoStop}%
\bibitem [{\citenamefont {Bersier}\ \emph {et~al.}(2009)\citenamefont
  {Bersier}, \citenamefont {Floris}, \citenamefont {Cudazzo}, \citenamefont
  {Profeta}, \citenamefont {Sanna}, \citenamefont {Bernardini}, \citenamefont
  {Monni}, \citenamefont {Pittalis}, \citenamefont {Sharma}, \citenamefont
  {Glawe}, \citenamefont {Continenza}, \citenamefont {Massidda},\ and\
  \citenamefont {Gross}}]{bers:09}%
  \BibitemOpen
  \bibfield  {author} {\bibinfo {author} {\bibfnamefont {C.}~\bibnamefont
  {Bersier}}, \bibinfo {author} {\bibfnamefont {A.}~\bibnamefont {Floris}},
  \bibinfo {author} {\bibfnamefont {P.}~\bibnamefont {Cudazzo}}, \bibinfo
  {author} {\bibfnamefont {G.}~\bibnamefont {Profeta}}, \bibinfo {author}
  {\bibfnamefont {A.}~\bibnamefont {Sanna}}, \bibinfo {author} {\bibfnamefont
  {F.}~\bibnamefont {Bernardini}}, \bibinfo {author} {\bibfnamefont
  {M.}~\bibnamefont {Monni}}, \bibinfo {author} {\bibfnamefont
  {S.}~\bibnamefont {Pittalis}}, \bibinfo {author} {\bibfnamefont
  {S.}~\bibnamefont {Sharma}}, \bibinfo {author} {\bibfnamefont
  {H.}~\bibnamefont {Glawe}}, \bibinfo {author} {\bibfnamefont
  {A.}~\bibnamefont {Continenza}}, \bibinfo {author} {\bibfnamefont
  {S.}~\bibnamefont {Massidda}},\ and\ \bibinfo {author} {\bibfnamefont
  {E.~K.~U.}\ \bibnamefont {Gross}},\ }\bibfield  {title} {\bibinfo {title}
  {{Multiband superconductivity in Pb, H under pressure and CaBeSi from
  ab-initio calculations}},\ }\href
  {https://doi.org/10.1088/0953-8984/21/16/164209} {\bibfield  {journal}
  {\bibinfo  {journal} {Journal of Physics: Condensed Matter}\ }\textbf
  {\bibinfo {volume} {21}},\ \bibinfo {pages} {164209} (\bibinfo {year}
  {2009})}\BibitemShut {NoStop}%
\bibitem [{\citenamefont {Wallace}(1972)}]{wall:72}%
  \BibitemOpen
  \bibfield  {author} {\bibinfo {author} {\bibfnamefont {D.~C.}\ \bibnamefont
  {Wallace}},\ }\bibfield  {title} {\bibinfo {title} {Thermodynamics of
  crystals},\ }\href {https://doi.org/10.1119/1.1987046} {\bibfield  {journal}
  {\bibinfo  {journal} {American Journal of Physics}\ }\textbf {\bibinfo
  {volume} {40}},\ \bibinfo {pages} {1718} (\bibinfo {year}
  {1972})}\BibitemShut {NoStop}%
\bibitem [{\citenamefont {Ruden}\ and\ \citenamefont {Wu}(1991)}]{rude:91}%
  \BibitemOpen
  \bibfield  {author} {\bibinfo {author} {\bibfnamefont {P.~P.}\ \bibnamefont
  {Ruden}}\ and\ \bibinfo {author} {\bibfnamefont {Z.}~\bibnamefont {Wu}},\
  }\bibfield  {title} {\bibinfo {title} {Exchange effect in coupled
  two‐dimensional electron gas systems},\ }\href
  {https://doi.org/10.1063/1.106116} {\bibfield  {journal} {\bibinfo  {journal}
  {Applied Physics Letters}\ }\textbf {\bibinfo {volume} {59}},\ \bibinfo
  {pages} {2165} (\bibinfo {year} {1991})}\BibitemShut {NoStop}%
\bibitem [{\citenamefont {Scopigno}\ \emph {et~al.}(2016)\citenamefont
  {Scopigno}, \citenamefont {Bucheli}, \citenamefont {Caprara}, \citenamefont
  {Biscaras}, \citenamefont {Bergeal}, \citenamefont {Lesueur},\ and\
  \citenamefont {Grilli}}]{Scop:16}%
  \BibitemOpen
  \bibfield  {author} {\bibinfo {author} {\bibfnamefont {N.}~\bibnamefont
  {Scopigno}}, \bibinfo {author} {\bibfnamefont {D.}~\bibnamefont {Bucheli}},
  \bibinfo {author} {\bibfnamefont {S.}~\bibnamefont {Caprara}}, \bibinfo
  {author} {\bibfnamefont {J.}~\bibnamefont {Biscaras}}, \bibinfo {author}
  {\bibfnamefont {N.}~\bibnamefont {Bergeal}}, \bibinfo {author} {\bibfnamefont
  {J.}~\bibnamefont {Lesueur}},\ and\ \bibinfo {author} {\bibfnamefont
  {M.}~\bibnamefont {Grilli}},\ }\bibfield  {title} {\bibinfo {title} {Phase
  separation from electron confinement at oxide interfaces},\ }\href
  {https://doi.org/10.1103/PhysRevLett.116.026804} {\bibfield  {journal}
  {\bibinfo  {journal} {Phys. Rev. Lett.}\ }\textbf {\bibinfo {volume} {116}},\
  \bibinfo {pages} {026804} (\bibinfo {year} {2016})}\BibitemShut {NoStop}%
\bibitem [{\citenamefont {Steffen}\ \emph {et~al.}(2017)\citenamefont
  {Steffen}, \citenamefont {Fr\'esard},\ and\ \citenamefont {Kopp}}]{stef:17}%
  \BibitemOpen
  \bibfield  {author} {\bibinfo {author} {\bibfnamefont {K.}~\bibnamefont
  {Steffen}}, \bibinfo {author} {\bibfnamefont {R.}~\bibnamefont {Fr\'esard}},\
  and\ \bibinfo {author} {\bibfnamefont {T.}~\bibnamefont {Kopp}},\ }\bibfield
  {title} {\bibinfo {title} {Capacitance and compressibility of
  heterostructures with strong electronic correlations},\ }\href
  {https://doi.org/10.1103/PhysRevB.95.035143} {\bibfield  {journal} {\bibinfo
  {journal} {Phys. Rev. B}\ }\textbf {\bibinfo {volume} {95}},\ \bibinfo
  {pages} {035143} (\bibinfo {year} {2017})}\BibitemShut {NoStop}%
\bibitem [{\citenamefont {Nagano}\ \emph {et~al.}(1984)\citenamefont {Nagano},
  \citenamefont {Singwi},\ and\ \citenamefont {Ohnishi}}]{naga:84}%
  \BibitemOpen
  \bibfield  {author} {\bibinfo {author} {\bibfnamefont {S.}~\bibnamefont
  {Nagano}}, \bibinfo {author} {\bibfnamefont {K.~S.}\ \bibnamefont {Singwi}},\
  and\ \bibinfo {author} {\bibfnamefont {S.}~\bibnamefont {Ohnishi}},\
  }\bibfield  {title} {\bibinfo {title} {Correlations in a two-dimensional
  quantum electron gas: The ladder approximation},\ }\href
  {https://doi.org/10.1103/PhysRevB.29.1209} {\bibfield  {journal} {\bibinfo
  {journal} {Phys. Rev. B}\ }\textbf {\bibinfo {volume} {29}},\ \bibinfo
  {pages} {1209} (\bibinfo {year} {1984})}\BibitemShut {NoStop}%
\bibitem [{\citenamefont {Santander-Syro}\ \emph {et~al.}(2011)\citenamefont
  {Santander-Syro}, \citenamefont {Copie}, \citenamefont {Kondo}, \citenamefont
  {Fortuna}, \citenamefont {Pailh{\`e}s}, \citenamefont {Weht}, \citenamefont
  {Qiu}, \citenamefont {Bertran}, \citenamefont {Nicolaou}, \citenamefont
  {Taleb-Ibrahimi}, \citenamefont {Le~F{\`e}vre}, \citenamefont {Herranz},
  \citenamefont {Bibes}, \citenamefont {Reyren}, \citenamefont {Apertet},
  \citenamefont {Lecoeur}, \citenamefont {Barth{\'e}l{\'e}my},\ and\
  \citenamefont {Rozenberg}}]{sant:11}%
  \BibitemOpen
  \bibfield  {author} {\bibinfo {author} {\bibfnamefont {A.~F.}\ \bibnamefont
  {Santander-Syro}}, \bibinfo {author} {\bibfnamefont {O.}~\bibnamefont
  {Copie}}, \bibinfo {author} {\bibfnamefont {T.}~\bibnamefont {Kondo}},
  \bibinfo {author} {\bibfnamefont {F.}~\bibnamefont {Fortuna}}, \bibinfo
  {author} {\bibfnamefont {S.}~\bibnamefont {Pailh{\`e}s}}, \bibinfo {author}
  {\bibfnamefont {R.}~\bibnamefont {Weht}}, \bibinfo {author} {\bibfnamefont
  {X.~G.}\ \bibnamefont {Qiu}}, \bibinfo {author} {\bibfnamefont
  {F.}~\bibnamefont {Bertran}}, \bibinfo {author} {\bibfnamefont
  {A.}~\bibnamefont {Nicolaou}}, \bibinfo {author} {\bibfnamefont
  {A.}~\bibnamefont {Taleb-Ibrahimi}}, \bibinfo {author} {\bibfnamefont
  {P.}~\bibnamefont {Le~F{\`e}vre}}, \bibinfo {author} {\bibfnamefont
  {G.}~\bibnamefont {Herranz}}, \bibinfo {author} {\bibfnamefont
  {M.}~\bibnamefont {Bibes}}, \bibinfo {author} {\bibfnamefont
  {N.}~\bibnamefont {Reyren}}, \bibinfo {author} {\bibfnamefont
  {Y.}~\bibnamefont {Apertet}}, \bibinfo {author} {\bibfnamefont
  {P.}~\bibnamefont {Lecoeur}}, \bibinfo {author} {\bibfnamefont
  {A.}~\bibnamefont {Barth{\'e}l{\'e}my}},\ and\ \bibinfo {author}
  {\bibfnamefont {M.~J.}\ \bibnamefont {Rozenberg}},\ }\bibfield  {title}
  {\bibinfo {title} {Two-dimensional electron gas with universal subbands at
  the surface of srtio3},\ }\href {https://doi.org/10.1038/nature09720}
  {\bibfield  {journal} {\bibinfo  {journal} {Nature}\ }\textbf {\bibinfo
  {volume} {469}},\ \bibinfo {pages} {189} (\bibinfo {year}
  {2011})}\BibitemShut {NoStop}%
\bibitem [{\citenamefont {Huijben}\ \emph {et~al.}(2009)\citenamefont
  {Huijben}, \citenamefont {Brinkman}, \citenamefont {Koster}, \citenamefont
  {Rijnders}, \citenamefont {Hilgenkamp},\ and\ \citenamefont
  {Blank}}]{huij:09}%
  \BibitemOpen
  \bibfield  {author} {\bibinfo {author} {\bibfnamefont {M.}~\bibnamefont
  {Huijben}}, \bibinfo {author} {\bibfnamefont {A.}~\bibnamefont {Brinkman}},
  \bibinfo {author} {\bibfnamefont {G.}~\bibnamefont {Koster}}, \bibinfo
  {author} {\bibfnamefont {G.}~\bibnamefont {Rijnders}}, \bibinfo {author}
  {\bibfnamefont {H.}~\bibnamefont {Hilgenkamp}},\ and\ \bibinfo {author}
  {\bibfnamefont {D.~H.}\ \bibnamefont {Blank}},\ }\bibfield  {title} {\bibinfo
  {title} {Structure--property relation of srtio$_3$/laalo$_3$ interfaces},\
  }\href@noop {} {\bibfield  {journal} {\bibinfo  {journal} {Advanced
  Materials}\ }\textbf {\bibinfo {volume} {21}},\ \bibinfo {pages} {1665}
  (\bibinfo {year} {2009})}\BibitemShut {NoStop}%
\bibitem [{\citenamefont {Gariglio}\ \emph {et~al.}(2009)\citenamefont
  {Gariglio}, \citenamefont {Reyren}, \citenamefont {Caviglia},\ and\
  \citenamefont {Triscone}}]{Gari:09}%
  \BibitemOpen
  \bibfield  {author} {\bibinfo {author} {\bibfnamefont {S.}~\bibnamefont
  {Gariglio}}, \bibinfo {author} {\bibfnamefont {N.}~\bibnamefont {Reyren}},
  \bibinfo {author} {\bibfnamefont {A.~D.}\ \bibnamefont {Caviglia}},\ and\
  \bibinfo {author} {\bibfnamefont {J.-M.}\ \bibnamefont {Triscone}},\
  }\bibfield  {title} {\bibinfo {title} {Superconductivity at the
  {LaAlO}$_3$/{SrTiO}$_3$ interface},\ }\href
  {https://doi.org/10.1088/0953-8984/21/16/164213} {\bibfield  {journal}
  {\bibinfo  {journal} {Journal of Physics: Condensed Matter}\ }\textbf
  {\bibinfo {volume} {21}},\ \bibinfo {pages} {164213} (\bibinfo {year}
  {2009})}\BibitemShut {NoStop}%
\bibitem [{\citenamefont {van~der Marel}\ \emph {et~al.}(2011)\citenamefont
  {van~der Marel}, \citenamefont {van Mechelen},\ and\ \citenamefont
  {Mazin}}]{vand:11}%
  \BibitemOpen
  \bibfield  {author} {\bibinfo {author} {\bibfnamefont {D.}~\bibnamefont
  {van~der Marel}}, \bibinfo {author} {\bibfnamefont {J.~L.~M.}\ \bibnamefont
  {van Mechelen}},\ and\ \bibinfo {author} {\bibfnamefont {I.~I.}\ \bibnamefont
  {Mazin}},\ }\bibfield  {title} {\bibinfo {title} {Common fermi-liquid origin
  of ${T}^{2}$ resistivity and superconductivity in $n$-type srtio$_{3}$},\
  }\href {https://doi.org/10.1103/PhysRevB.84.205111} {\bibfield  {journal}
  {\bibinfo  {journal} {Phys. Rev. B}\ }\textbf {\bibinfo {volume} {84}},\
  \bibinfo {pages} {205111} (\bibinfo {year} {2011})}\BibitemShut {NoStop}%
\bibitem [{\citenamefont {Gariglio}\ \emph {et~al.}(2015)\citenamefont
  {Gariglio}, \citenamefont {F{\^e}te},\ and\ \citenamefont
  {Triscone}}]{gari:15}%
  \BibitemOpen
  \bibfield  {author} {\bibinfo {author} {\bibfnamefont {S.}~\bibnamefont
  {Gariglio}}, \bibinfo {author} {\bibfnamefont {A.}~\bibnamefont {F{\^e}te}},\
  and\ \bibinfo {author} {\bibfnamefont {J.-M.}\ \bibnamefont {Triscone}},\
  }\bibfield  {title} {\bibinfo {title} {Electron confinement at the
  laalo3/srtio3 interface},\ }\href@noop {} {\bibfield  {journal} {\bibinfo
  {journal} {Journal of Physics: Condensed Matter}\ }\textbf {\bibinfo {volume}
  {27}},\ \bibinfo {pages} {283201} (\bibinfo {year} {2015})}\BibitemShut
  {NoStop}%
\bibitem [{\citenamefont {Bark}\ \emph {et~al.}(2011)\citenamefont {Bark},
  \citenamefont {Felker}, \citenamefont {Wang}, \citenamefont {Zhang},
  \citenamefont {Jang}, \citenamefont {Folkman}, \citenamefont {Park},
  \citenamefont {Baek}, \citenamefont {Zhou}, \citenamefont {Fong} \emph
  {et~al.}}]{bark:11}%
  \BibitemOpen
  \bibfield  {author} {\bibinfo {author} {\bibfnamefont {C.}~\bibnamefont
  {Bark}}, \bibinfo {author} {\bibfnamefont {D.}~\bibnamefont {Felker}},
  \bibinfo {author} {\bibfnamefont {Y.}~\bibnamefont {Wang}}, \bibinfo {author}
  {\bibfnamefont {Y.}~\bibnamefont {Zhang}}, \bibinfo {author} {\bibfnamefont
  {H.}~\bibnamefont {Jang}}, \bibinfo {author} {\bibfnamefont {C.}~\bibnamefont
  {Folkman}}, \bibinfo {author} {\bibfnamefont {J.}~\bibnamefont {Park}},
  \bibinfo {author} {\bibfnamefont {S.}~\bibnamefont {Baek}}, \bibinfo {author}
  {\bibfnamefont {H.}~\bibnamefont {Zhou}}, \bibinfo {author} {\bibfnamefont
  {D.}~\bibnamefont {Fong}}, \emph {et~al.},\ }\bibfield  {title} {\bibinfo
  {title} {{Tailoring a two-dimensional electron gas at the LaAlO$_3$/SrTiO$_3$
  (001) interface by epitaxial strain}},\ }\href@noop {} {\bibfield  {journal}
  {\bibinfo  {journal} {Proceedings of the National Academy of Sciences}\
  }\textbf {\bibinfo {volume} {108}},\ \bibinfo {pages} {4720} (\bibinfo {year}
  {2011})}\BibitemShut {NoStop}%
\bibitem [{\citenamefont {Neville}\ \emph {et~al.}(1972)\citenamefont
  {Neville}, \citenamefont {Hoeneisen},\ and\ \citenamefont {Mead}}]{nevi:72}%
  \BibitemOpen
  \bibfield  {author} {\bibinfo {author} {\bibfnamefont {R.~C.}\ \bibnamefont
  {Neville}}, \bibinfo {author} {\bibfnamefont {B.}~\bibnamefont {Hoeneisen}},\
  and\ \bibinfo {author} {\bibfnamefont {C.~A.}\ \bibnamefont {Mead}},\
  }\bibfield  {title} {\bibinfo {title} {Permittivity of strontium titanate},\
  }\href {https://doi.org/10.1063/1.1661463} {\bibfield  {journal} {\bibinfo
  {journal} {Journal of Applied Physics}\ }\textbf {\bibinfo {volume} {43}},\
  \bibinfo {pages} {2124} (\bibinfo {year} {1972})}\BibitemShut {NoStop}%
\bibitem [{\citenamefont {Berg}\ \emph {et~al.}(1995)\citenamefont {Berg},
  \citenamefont {Blom}, \citenamefont {Cillessen},\ and\ \citenamefont
  {Wolf}}]{berg:95}%
  \BibitemOpen
  \bibfield  {author} {\bibinfo {author} {\bibfnamefont {R.}~\bibnamefont
  {Berg}}, \bibinfo {author} {\bibfnamefont {P.}~\bibnamefont {Blom}}, \bibinfo
  {author} {\bibfnamefont {J.}~\bibnamefont {Cillessen}},\ and\ \bibinfo
  {author} {\bibfnamefont {R.}~\bibnamefont {Wolf}},\ }\bibfield  {title}
  {\bibinfo {title} {{Field dependent permittivity in metal‐semiconducting
  SrTiO$_3$ Schottky diodes}},\ }\href {https://doi.org/10.1063/1.114103}
  {\bibfield  {journal} {\bibinfo  {journal} {Applied Physics Letters}\
  }\textbf {\bibinfo {volume} {66}},\ \bibinfo {pages} {697} (\bibinfo {year}
  {1995})}\BibitemShut {NoStop}%
\bibitem [{\citenamefont {Sing}\ \emph {et~al.}(2009)\citenamefont {Sing},
  \citenamefont {Berner}, \citenamefont {Goss}, \citenamefont {Müller},
  \citenamefont {Ruff}, \citenamefont {Wetscherek}, \citenamefont {Paetel},
  \citenamefont {Mannhart}, \citenamefont {Pauli}, \citenamefont {Schneider},
  \citenamefont {Willmott}, \citenamefont {Gorgoi}, \citenamefont {Schäfers},\
  and\ \citenamefont {Claessen}}]{sing:09}%
  \BibitemOpen
  \bibfield  {author} {\bibinfo {author} {\bibfnamefont {M.}~\bibnamefont
  {Sing}}, \bibinfo {author} {\bibfnamefont {G.}~\bibnamefont {Berner}},
  \bibinfo {author} {\bibfnamefont {K.}~\bibnamefont {Goss}}, \bibinfo {author}
  {\bibfnamefont {A.}~\bibnamefont {Müller}}, \bibinfo {author} {\bibfnamefont
  {A.}~\bibnamefont {Ruff}}, \bibinfo {author} {\bibfnamefont {A.}~\bibnamefont
  {Wetscherek}}, \bibinfo {author} {\bibfnamefont {S.}~\bibnamefont {Paetel}},
  \bibinfo {author} {\bibfnamefont {J.}~\bibnamefont {Mannhart}}, \bibinfo
  {author} {\bibfnamefont {S.}~\bibnamefont {Pauli}}, \bibinfo {author}
  {\bibfnamefont {C.}~\bibnamefont {Schneider}}, \bibinfo {author}
  {\bibfnamefont {P.}~\bibnamefont {Willmott}}, \bibinfo {author}
  {\bibfnamefont {M.}~\bibnamefont {Gorgoi}}, \bibinfo {author} {\bibfnamefont
  {F.}~\bibnamefont {Schäfers}},\ and\ \bibinfo {author} {\bibfnamefont
  {R.}~\bibnamefont {Claessen}},\ }\bibfield  {title} {\bibinfo {title}
  {{Profiling the Interface Electron Gas of LaAlO$_3$/SrTiO$_3$
  Heterostructures with Hard X-Ray Photoelectron Spectroscopy}},\ }\href
  {https://doi.org/10.1103/PhysRevLett.102.176805} {\bibfield  {journal}
  {\bibinfo  {journal} {Physical Review Letters}\ }\textbf {\bibinfo {volume}
  {102}},\ \bibinfo {pages} {176805} (\bibinfo {year} {2009})}\BibitemShut
  {NoStop}%
\bibitem [{\citenamefont {Janotti}\ \emph {et~al.}(2011)\citenamefont
  {Janotti}, \citenamefont {Steiauf},\ and\ \citenamefont {Walle}}]{jano:11}%
  \BibitemOpen
  \bibfield  {author} {\bibinfo {author} {\bibfnamefont {A.}~\bibnamefont
  {Janotti}}, \bibinfo {author} {\bibfnamefont {D.}~\bibnamefont {Steiauf}},\
  and\ \bibinfo {author} {\bibfnamefont {C.}~\bibnamefont {Walle}},\ }\bibfield
   {title} {\bibinfo {title} {{Strain effects on the electronic structure of
  SrTiO$_3$: Toward high electron mobilities}},\ }\href
  {https://doi.org/10.1103/PhysRevB.84.201304} {\bibfield  {journal} {\bibinfo
  {journal} {Physical Review B}\ }\textbf {\bibinfo {volume} {84}},\ \bibinfo
  {pages} {201304} (\bibinfo {year} {2011})}\BibitemShut {NoStop}%
\bibitem [{\citenamefont {Peelaers}\ \emph {et~al.}(2015)\citenamefont
  {Peelaers}, \citenamefont {Krishnaswamy}, \citenamefont {Gordon},
  \citenamefont {Steiauf}, \citenamefont {Sarwe}, \citenamefont {Janotti},\
  and\ \citenamefont {Van~de Walle}}]{peel:15}%
  \BibitemOpen
  \bibfield  {author} {\bibinfo {author} {\bibfnamefont {H.}~\bibnamefont
  {Peelaers}}, \bibinfo {author} {\bibfnamefont {K.}~\bibnamefont
  {Krishnaswamy}}, \bibinfo {author} {\bibfnamefont {L.}~\bibnamefont
  {Gordon}}, \bibinfo {author} {\bibfnamefont {D.}~\bibnamefont {Steiauf}},
  \bibinfo {author} {\bibfnamefont {A.}~\bibnamefont {Sarwe}}, \bibinfo
  {author} {\bibfnamefont {A.}~\bibnamefont {Janotti}},\ and\ \bibinfo {author}
  {\bibfnamefont {C.}~\bibnamefont {Van~de Walle}},\ }\bibfield  {title}
  {\bibinfo {title} {Impact of electric-field dependent dielectric constants on
  two-dimensional electron gases in complex oxides},\ }\href
  {https://doi.org/10.1063/1.4935222} {\bibfield  {journal} {\bibinfo
  {journal} {Applied Physics Letters}\ }\textbf {\bibinfo {volume} {107}},\
  \bibinfo {pages} {183505} (\bibinfo {year} {2015})}\BibitemShut {NoStop}%
\bibitem [{\citenamefont {Copie}\ \emph {et~al.}(2009)\citenamefont {Copie},
  \citenamefont {Garcia}, \citenamefont {Bödefeld}, \citenamefont
  {Carrétéro}, \citenamefont {Bibes}, \citenamefont {Herranz}, \citenamefont
  {Jacquet}, \citenamefont {Maurice}, \citenamefont {Vinter}, \citenamefont
  {Fusil}, \citenamefont {Bouzehouane}, \citenamefont {Jaffrès},\ and\
  \citenamefont {Barthelemy}}]{copi:09}%
  \BibitemOpen
  \bibfield  {author} {\bibinfo {author} {\bibfnamefont {O.}~\bibnamefont
  {Copie}}, \bibinfo {author} {\bibfnamefont {V.}~\bibnamefont {Garcia}},
  \bibinfo {author} {\bibfnamefont {C.}~\bibnamefont {Bödefeld}}, \bibinfo
  {author} {\bibfnamefont {C.}~\bibnamefont {Carrétéro}}, \bibinfo {author}
  {\bibfnamefont {M.}~\bibnamefont {Bibes}}, \bibinfo {author} {\bibfnamefont
  {G.}~\bibnamefont {Herranz}}, \bibinfo {author} {\bibfnamefont
  {E.}~\bibnamefont {Jacquet}}, \bibinfo {author} {\bibfnamefont {J.-L.}\
  \bibnamefont {Maurice}}, \bibinfo {author} {\bibfnamefont {B.}~\bibnamefont
  {Vinter}}, \bibinfo {author} {\bibfnamefont {S.}~\bibnamefont {Fusil}},
  \bibinfo {author} {\bibfnamefont {K.}~\bibnamefont {Bouzehouane}}, \bibinfo
  {author} {\bibfnamefont {H.}~\bibnamefont {Jaffrès}},\ and\ \bibinfo
  {author} {\bibfnamefont {A.}~\bibnamefont {Barthelemy}},\ }\bibfield  {title}
  {\bibinfo {title} {{Towards Two-Dimensional Metallic Behavior at
  LaAlO$_3$/SrTiO$_3$ Interfaces}},\ }\href
  {https://doi.org/10.1103/PhysRevLett.102.216804} {\bibfield  {journal}
  {\bibinfo  {journal} {Physical Review Letters}\ }\textbf {\bibinfo {volume}
  {102}},\ \bibinfo {pages} {216804} (\bibinfo {year} {2009})}\BibitemShut
  {NoStop}%
\bibitem [{\citenamefont {Maznichenko}\ \emph {et~al.}(2018)\citenamefont
  {Maznichenko}, \citenamefont {Ostanin}, \citenamefont {Dugaev}, \citenamefont
  {Mertig},\ and\ \citenamefont {Ernst}}]{mazn:18}%
  \BibitemOpen
  \bibfield  {author} {\bibinfo {author} {\bibfnamefont {I.}~\bibnamefont
  {Maznichenko}}, \bibinfo {author} {\bibfnamefont {S.}~\bibnamefont
  {Ostanin}}, \bibinfo {author} {\bibfnamefont {V.}~\bibnamefont {Dugaev}},
  \bibinfo {author} {\bibfnamefont {I.}~\bibnamefont {Mertig}},\ and\ \bibinfo
  {author} {\bibfnamefont {A.}~\bibnamefont {Ernst}},\ }\bibfield  {title}
  {\bibinfo {title} {{Impact of long-range disorder on the two-dimensional
  electron gas formation at a LaAlO$_3$ / SrTiO$_3$ interface}},\ }\href@noop
  {} {\bibfield  {journal} {\bibinfo  {journal} {Physical Review Materials}\
  }\textbf {\bibinfo {volume} {2}},\ \bibinfo {pages} {074003} (\bibinfo {year}
  {2018})}\BibitemShut {NoStop}%
\bibitem [{\citenamefont {Guo}\ \emph {et~al.}(2016)\citenamefont {Guo},
  \citenamefont {Saidi},\ and\ \citenamefont {Zhao}}]{Guo:16}%
  \BibitemOpen
  \bibfield  {author} {\bibinfo {author} {\bibfnamefont {H.}~\bibnamefont
  {Guo}}, \bibinfo {author} {\bibfnamefont {W.~A.}\ \bibnamefont {Saidi}},\
  and\ \bibinfo {author} {\bibfnamefont {J.}~\bibnamefont {Zhao}},\ }\bibfield
  {title} {\bibinfo {title} {{Tunability of the Two-Dimensional Electron Gas at
  LaAlO$_3$/SrTiO$_3$ Interface by strain-Induced Ferroelectricity}},\
  }\href@noop {} {\bibfield  {journal} {\bibinfo  {journal} {Physical Chemistry
  Chemical Physics}\ }\textbf {\bibinfo {volume} {18}},\ \bibinfo {pages}
  {28474} (\bibinfo {year} {2016})}\BibitemShut {NoStop}%
\bibitem [{\citenamefont {Zhong}\ \emph {et~al.}(2013)\citenamefont {Zhong},
  \citenamefont {T\'oth},\ and\ \citenamefont {Held}}]{zhon:13}%
  \BibitemOpen
  \bibfield  {author} {\bibinfo {author} {\bibfnamefont {Z.}~\bibnamefont
  {Zhong}}, \bibinfo {author} {\bibfnamefont {A.}~\bibnamefont {T\'oth}},\ and\
  \bibinfo {author} {\bibfnamefont {K.}~\bibnamefont {Held}},\ }\bibfield
  {title} {\bibinfo {title} {{Theory of spin-orbit coupling at
  LaAlO$_{3}$/SrTiO$_{3}$ interfaces and SrTiO$_{3}$ surfaces}},\ }\href
  {https://doi.org/10.1103/PhysRevB.87.161102} {\bibfield  {journal} {\bibinfo
  {journal} {Phys. Rev. B}\ }\textbf {\bibinfo {volume} {87}},\ \bibinfo
  {pages} {161102} (\bibinfo {year} {2013})}\BibitemShut {NoStop}%
\bibitem [{\citenamefont {McCollam}\ \emph {et~al.}(2014)\citenamefont
  {McCollam}, \citenamefont {Wenderich}, \citenamefont {Kruize}, \citenamefont
  {Guduru}, \citenamefont {Molegraaf}, \citenamefont {Huijben}, \citenamefont
  {Koster}, \citenamefont {Blank}, \citenamefont {Rijnders}, \citenamefont
  {Brinkman} \emph {et~al.}}]{mcco:14}%
  \BibitemOpen
  \bibfield  {author} {\bibinfo {author} {\bibfnamefont {A.}~\bibnamefont
  {McCollam}}, \bibinfo {author} {\bibfnamefont {S.}~\bibnamefont {Wenderich}},
  \bibinfo {author} {\bibfnamefont {M.}~\bibnamefont {Kruize}}, \bibinfo
  {author} {\bibfnamefont {V.}~\bibnamefont {Guduru}}, \bibinfo {author}
  {\bibfnamefont {H.}~\bibnamefont {Molegraaf}}, \bibinfo {author}
  {\bibfnamefont {M.}~\bibnamefont {Huijben}}, \bibinfo {author} {\bibfnamefont
  {G.}~\bibnamefont {Koster}}, \bibinfo {author} {\bibfnamefont {D.~H.}\
  \bibnamefont {Blank}}, \bibinfo {author} {\bibfnamefont {G.}~\bibnamefont
  {Rijnders}}, \bibinfo {author} {\bibfnamefont {A.}~\bibnamefont {Brinkman}},
  \emph {et~al.},\ }\bibfield  {title} {\bibinfo {title} {{Quantum oscillations
  and subband properties of the two-dimensional electron gas at the
  LaAlO$_3$/SrTiO$_3$ interface}},\ }\href@noop {} {\bibfield  {journal}
  {\bibinfo  {journal} {APL materials}\ }\textbf {\bibinfo {volume} {2}},\
  \bibinfo {pages} {022102} (\bibinfo {year} {2014})}\BibitemShut {NoStop}%
\end{thebibliography}%

\end{document}